\documentclass[12pt]{article}

\usepackage{times}
\usepackage[greek,english]{babel}
\usepackage{graphicx}
\usepackage[figurename=Figure,font=footnotesize,labelfont=bf]{caption}
\usepackage{color}
\usepackage{multicol}
\usepackage{multirow}
\usepackage{url}
\usepackage{amsmath}
\usepackage[noadjust]{cite}
\usepackage{pdfpages}

\title{General theory for packing icosahedral shells into multi-component aggregates}

\author
{Nicolò Canestrari$^1$, Diana Nelli,$^{1\ast}$ Riccardo Ferrando$^{1\ast}$\\
\\
\normalsize{1 Dipartimento di Fisica, Università di Genova, Via Dodecaneso 33, 16146 Genova, Italy}\\
\\
\normalsize{$^\ast$To whom correspondence should be addressed;}\\
\normalsize{E-mail: diana.nelli@edu.unige.it; riccardo.ferrando@unige.it.}
}

\date{}

\begin{document} 


\baselineskip24pt


\maketitle


\section*{Abstract}
Multi-component aggregates are being intensively researched in various fields because of their highly tunable properties and wide applications. Due to the complex configurational space of these systems, research would greatly benefit from a general theoretical framework for the prediction of stable structures, which, however, is largely incomplete at present. Here we propose a general theory for the construction of multi-component icosahedral structures by assembling concentric shells of different chiral and achiral types, consisting of particles of different sizes. By mapping shell sequences into paths in the hexagonal lattice, we establish simple and general rules for designing a wide variety of magic icosahedral structures, and we evaluate the optimal size-mismatch between particles in the different shells. The predictions of our design strategy are confirmed by molecular dynamics simulations and density functional theory calculations for several multi-component atomic clusters and nanoparticles.

\section*{Introduction}
The research on multi-components nanoaggregates is very active and spans many different fields. High-entropy alloy nanocrystals, consisting of nanometer-sized solid solutions of five or more elements, have attracted much attention due to their enhanced structural stability and catalytic activity \cite{Yao2022science,Yao2020sciadv,Xie2019natcomm,Yan2022natcomm}. Ordered architectures are explored as well. Among them, the assembly of concentric shells of different compositions into multilayer aggregates is a widely employed tool to protect or functionalize the core \cite{Howes2014science,Chen2021natcomm}, improve the stability \cite{Wang2013natmat,Huo2006jacs,Vega2023acsnano,Foucher2023chemmat}, and adjust the surface properties \cite{Nelli2023apx,Strasser2010nc} of the nanostructure.

The wide range of possible compositions and the huge configurational space of multi-component systems are key to their broad success. However, such inherent complexity poses major challenges to the design and synthesis of nanoaggregates with well-defined and durable configurations. A general theory for the prediction of stable multicomponent structures would be an essential reference for the design and synthesis of nanoparticles for customised applications. However, such a theory is lacking at present.

Here we propose a theoretical approach which generalizes and unify concepts from crystallography \cite{Mackay1962ac} and structural biology \cite{Caspar1962cshs,SadreMarandi2018cmb,Twarock2019ncomms} to develop a design strategy of multi-component clusters and nanoparticles. 

We consider multi-component aggregates formed by particles of different sizes, and establish general criteria for assembling these particles into highly symmetric multi-shell structures. In particular, we apply our approach to  icosahedral structures.    
Icosahedra combine the maximum symmetry with the most compact shape. These properties favour energy stability in clusters and nanoparticles \cite{Martin1996pr,Baletto2005rmp} and give an evolutionary advantage in biological systems such as viruses \cite{Caspar1962cshs,Twarock2019ncomms,Li2018pnas}. Accordingly, icosahedra have been observed in a huge variety of systems, including clusters and nanoparticles \cite{Farges1981ss, Wang2012prl,Langille2012science,Hubert1998nature,Nelli2023nanoscale}, colloidal aggregates \cite{deNijs2015natphys,Wang2018ncomms,Chen2021natphys}, intermetallic compounds and quasicrystals \cite{Shetchman1984prl,Noya2021nature,Pankova2015ic}, viral capsids, bacterial organelles, DNA and protein aggregates \cite{Rossmann1989arb,Salunke1989bj,Yeates2011cosb,Zandi2020pr,Douglas2009nature}. 

Several research efforts can be found in the literature, starting with the seminal work of Caspar and Klug \cite{Caspar1962cshs}, dedicated to rationalising the structure of individual icosahedral shells, especially in the field of virus biology \cite{SadreMarandi2018cmb,Twarock2019ncomms,MartinBravo2021acsnano,Pinto2023pnas}. Icosahedral shells are made of one layer of particles, which can be arranged according to achiral or chiral symmetries. Chiral shells present all rotational symmetry operations of the icosahedron, but lack its reflection planes.

On the contrary, there is no general theory for assembling together multiple shells of different radii, with achiral and chiral symmetries, into concentric arrangements to generate compact icosahedral aggregates. In metal nanoparticles and clusters, and in aggregates of colloidal particles, compact structures are much more commonly observed than single shells \cite{Farges1981ss,Martin1996pr,Wang2012prl,Nelli2023nanoscale,deNijs2015natphys,Wang2018ncomms,Chen2021natphys}. The theoretical efforts to assemble multi-shell icosahedra are rather limited. They began with the historical works of Bergman et al. \cite{Bergman1957ac} and Mackay \cite{Mackay1962ac}, both concerning achiral structures only. In particular, Mackay constructed icosahedra by packing spheres of equal size arranged in shells around a central particle. Mackay icosahedra turns out to be built of 20 distorted tetrahedra, in which particle layers are arranged into the stacking of the face-centered cubic lattice (ABCABC...). Most icosahedra observed in metal clusters and nanoparticles are of Mackay type \cite{Farges1981ss,Wang2012prl,Langille2012science,Nelli2023nanoscale}. Mackay also proposed a possible termination by a single shell in hexagonal close-packed (hcp) stacking (giving for example the sequence ABCABCB), known as the anti-Mackay shell. Anti-Mackay terminations have been recently observed in colloidal aggregates \cite{deNijs2015natphys,Wang2018ncomms}. Even fewer theoretical efforts have been devoted to construct icosahedra that include chiral shells. These efforts have been limited to adding a single chiral shell on top of an achiral Mackay core \cite{Bochicchio2010nl,Settem2020cms,Whetten2019acr}.  


The key point of our theoretical approach is the mapping of multi-shell structures into paths in the hexagonal lattice. The mapping naturally leads to the design of multi-shell sequences corresponding to alternative series of icosahedral magic numbers. Furthermore, the mapping allows us to predict which sequences exhibit spontaneous symmetry breaking from achiral to chiral structures. In the field of metal clusters and nanoparticles, chiral icosahedra are especially interesting for their applications to catalysis \cite{Whetten2019acr}. 
In all cases, we demonstrate that icosahedra are stabilized by the size mismatch between particles of different shells and evaluate the optimal mismatch for energetic stability.
We note that, due to the contraction of pair distances between particles in adjacent shells \cite{Mackay1962ac}, the icosahedron is naturally suitable for accommodating particles of different sizes in different shells. Our approach is generalized also to anti-Mackay shells, so that it allows us to derive a design principle for the structures observed in colloidal aggregates \cite{deNijs2015natphys,Wang2018ncomms,Chen2021natphys}, and to predict others.

Our design strategy is validated by numerical calculations for several model systems and by ab-initio calculations for alkali metal clusters. The predictions of our theory are further confirmed by simulations of the growth of alkali and transition metal nanoparticles including up to four different elements. These simulations show that atoms naturally self-assemble into the predicted multi-shell icosahedral structures, including those with symmetry-breaking shell  sequences. 
Although the applications presented below concern atomic clusters and nanoparticles, due to the general character of its basic assumptions, our theory can be applied to the design of aggregates of other particle types, such aggregates of colloidal particles and complex molecules of biological interest.

\section*{Results}

\subsection*{Mapping icosahedra into paths}

Here we develop our theoretical framework, whose starting point of is the well-known approach of Caspar and Klug (CK) \cite{Caspar1962cshs,Twarock2019ncomms,SadreMarandi2018cmb}, which was originally proposed for rationalizing and predicting the architecture of icosahedral viral capsids. Specifically, they developped a general method for the construction of individual icosahedral shells, which is based on cutting and folding leaflets from the two-dimensional hexagonal lattice and produces achiral and chiral arrangements of particles on the icosahedral surface.

The CK construction is shown in Figure 1a. A segment is drawn between points of coordinates $(0,0)$ and $(h,k)$ with respect to the basis vectors of the hexagonal lattice; $h$ and $k$ are integer non-negative numbers, so that the segment always connects two lattice points. The segment is the base of an equilateral triangle, which is replicated 20 times to form a leaflet, which is then cut and folded to generate an icosahedral shell with a well-defined surface lattice.

In the CK theory, the triangulation number $T$ of an icosahedral shell is defined as the square length of the triangular edge, and is calculated as $T = h^2 + k^2 + hk$. Edge and radius of the shell are $\sqrt{T}$ and  $ \sin (2\textgreek{π} /5) \sqrt{T}$, respectively. Assigning one particle to each lattice point, the shell contains $10 \,T+2$ particles (see Supplementary Note 1.1).

Shells are achiral or chiral depending on the angle $\theta_\text{CK}$ between the $h$-axis and the segment of the CK construction. Achiral shells correspond to segments with $\theta_\text{CK} = 0^\circ, 30^\circ$ and 60$^\circ$. Segments on a coordinate axis correspond to the achiral shells described by Mackay (MC) in his work on the packing of equal spheres \cite{Mackay1962ac}. Shells built on segments on the diagonal ($h=k$, $\theta_\text{CK} = 30^\circ$) are here called of Bergman (BG) type, since the smallest is the outer shell of the Bergman cluster \cite{Pankova2015ic,Bergman1957ac}. All other shells are chiral, with enantiomers  symmetrically placed with respect to the diagonal. We remark that, in the CK theory, an icosahedral shell is uniquely determined by the segment endpoint $(h,k)$ in the hexagonal lattice; therefore, in the following, icosahedral shells will be denoted by their $(h,k)$.

We begin our generalised construction by grouping icosahedral shells into chirality classes. In Figure 1b, points in the hexagonal lattice are coloured according to the chirality class of the corresponding icosahedral shell. From each lattice point on the diagonal a chirality class originates, which we call Ch$n$; this comprises the $(n,n)$ BG shell, the shells with $h=n$ and $k>n$, and their enantiomers $(h>n,k=n)$. For example, shells of the Ch1 class have either $h=1$ or $k=1$, whereas the second index increases starting from $1$. MC shells are grouped into class Ch$0$, together with the one-particle shell $(0,0)$. The radius and the number of particles in the shell increase with the non-constant index, so that larger and larger shells are found while moving farther from the diagonal. Shells within the same chirality class share a similar particle arrangement on the icosahedral surface, whereas shells belonging to different classes are clearly different (see Figure 1c,d). The grouping of icosahedral shells into chirality classes is key to rationalize and predict their optimal packing, and their dynamic assembling into tightly packed structures. This will be clarified and deeply discussed in the following. 

The main point of our construction consists in assembling concentric shells into aggregates by drawing paths in the hexagonal lattice. As a first example, we consider a path along a coordinate axis, e.g., the $k$ axis (Figure 2a), starting from $k=0$ and making steps $(0,k)\to (0,k+1)$ up to $k=i-1$. This path assembles $i$ concentric MC shells of larger and larger size into a Mackay icosahedron \cite{Mackay1962ac,Martin1996pr}, a well-known structure observed in many experiments on clusters \cite{Farges1981ss,Wang2012prl,Nelli2023nanoscale,Branz2002prb,Koga2003ss}, which is thus recovered as a special case of our construction. Mackay icosahedra are tightly packed structures consisting of 20 distorted tetrahedra \cite{Martin1996pr} in which the particles are arranged according to the face-centered-cubic (fcc) lattice. The numbers of particles in a MC icosahedron made of $i$ shells is $N_i = (10 i^3 - 15 i^2 + 11i - 3)/3$, which gives the series of magic numbers $1$, $13$, $55$, $147$, $309$,...  

In general, paths can be drawn by allowing different increments at each step. Here we deal with the simplest generalization, which consists of choosing between $(h,k) \to (h,k+1)$ and $(h,k) \to (h+1,k)$ (other types of steps in the hexagonal plane will be discussed in the following). In this case, the number of shells in a path from the origin to a point $(h,k)$ is $i=h+k+1$. Such paths are inspired by the Mackay path of Figure 2a, in which one index is incremented at steps of one, whereas the other is kept constant and equal to zero. Allowing for different elementary steps in the hexagonal plane allows to build a wide variety of icosahedral structures, which retain the densely-packed character of the Mackay icosahedron.

We distinguish three cases, as shown in Figure 2b. If $(h,k)$ is above the diagonal, the elementary move $(h,k) \to (h,k+1)$ conserves the chirality class of the shell, i.e. the shell we are adding belongs to the same class of the previous one. On the contrary, by the move $(h,k) \to (h+1,k)$ the chirality class is incremented. If $(h,k)$ is below the diagonal, the opposite applies. For $(h,k)$ in the diagonal, i.e. for shells of BG type, both steps conserve the class; specifically, the two steps are equivalent since the corresponding added shells are enantiomers, with the same size and particle arrangement but opposite chirality. 

When assembling shells in physical systems, one must bear in mind that in the icosahedron the radius is shorter than the edge by $\sin (2\textgreek{π} /5) \approx 0.9511$, which has a direct effect on the packing of concentric shells. In the Mackay icosahedron of equal spheres \cite{Mackay1962ac}, the distance between spheres in neighbouring shells is shorter by about $5\%$ than that between spheres in the same shell. Similar considerations hold for shells belonging to other chirality classes, assembled according to the path rules identified so far; in some cases, the difference between intra-shell and inter-shell nearest-neighbour distances is even larger.
These considerations naturally lead to the idea of assembling shells in which particles in different shells have different sizes, which we better clarify below.

We consider the case of a core with $i$ shells, to which we add the outer shell $i+1$. Particles in the core and in the outer shell have different sizes. We define the size mismatch $\text{sm}_{i,i+1}=(d_{i+1}-d_i)/d_i$, with $d_{i+1}$ and $d_i$ particle sizes in shells $i+1$ and $i$, respectively.

In the example of Figure 3a-b, the core is made of five MC shells, i.e. it is terminated by the $(0,4)$ shell. According to our path rules, the outer shell can be either the $(0,5)$ MC shell (same class) or the $(1,4)$ Ch1 shell (different class). We consider a simple model, in which all particles interact by the well-known Lennard-Jones (LJ) potential; specifically, the interaction energy is the same for all particles, the only difference being the equilibrium distance of pair interactions, which accounts for particles of different sizes (see the Methods section). In Figure 2a, we calculate the binding energy depending on the size mismatch, and determine the mismatch that minimizes the energy of the whole aggregate. For both MC and Ch1 shells this optimal mismatch is positive, i.e. it is favourable to have bigger particles in the outer shell than in the core, and it is larger for Ch1 than for MC shells.

The optimal mismatch can be estimated also by geometric packing arguments, which are discussed in details in  Supplementary Note 1. Here we recall only the main assumption and give the final result. In order to evaluate the optimal mismatch between particles of shells $i$ and $i+1$ it is reasonable to impose that 
\begin{equation}
    r_{i+1} - r_i = \frac{d_{i}+d_{i+1}}{2}
\end{equation}
where $r_i$ and $r_{i+1}$ are the radii of the respective shells, $d_i$ and $d_{i+1}$ are the particle diameters in these shells. For atomic systems, these diameters are estimated from the nearest-neighbour distance of atoms in their crystal lattice. Recalling that the relation between the radii and the triangulation numbers of the shells
\begin{equation}
    r_i=  d_i \sin \left (\frac{2\textgreek{π}}{5}\right) \sqrt{T_i} 
\end{equation}
one finally obtains the following approximate expression for the optimal size mismatch 
\begin{equation}
    \text{sm}_{i,i+1} = \frac{2 \sin \left (\frac{2\textgreek{π}}{5}\right)(1+\xi) \sqrt{T_i}+1}{2 \sin \left (\frac{2\textgreek{π}}{5}\right)\sqrt{T_{i+1}}-1} - 1.
    \label{eq:sm}
\end{equation}
Here $\xi \ll 1$ is an expansion coefficient of pair distances in the core, which depends on the number and the type of shells of the same component in the core (see Supplementary Note 1.4).

In Figure 3c,d the results of Eq. (3) are compared to those for LJ and Morse clusters. Data in Figure 3c are obtained as in Figure 3a, but for different sizes of the Mackay core, whereas in Figure 3d we consider a cluster made of a Mackay core plus a Ch1 shell, to which we add a further shell of either Ch1 or Ch2 type. In all cases the agreement is good. Eq. (3) is thus a reliable guide for the semi-quantitative evaluation of the optimal mismatch, which demonstrates the key role of geometric factors in determining it.

For the paths of Figure 3, the optimal mismatch is always positive, i.e. icosahedral aggregates benefit from having bigger particles in the outer shells; in addition, much larger mismatches are found for class-changing than for class-conserving steps. These features are general for all icosahedra constructed according to the path rules identified so far, and are key to design stable icosahedral aggregates and to predict their natural growth modes.

\subsection*{Design strategy for icosahedral aggregates}

Our theory is now applied to the path-based design of icosahedral aggregates. In order to construct a multi-shell icosahedral aggregate, a path is drawn according to the rules in Figure 2 and, for each step, the optimal mismatch is estimated by Eq. (3). Then, particles of the appropriate sizes are associated with each shell.

An example is shown in Figure 4, where the path connecting BG shells through neighbouring chiral shells is considered (Figure 4a). This path alternates class-conserving and class-changing steps, spontaneously breaking mirror symmetries after the first BG shell. It produces the series of magic numbers
\begin{equation}
N_i= \frac{1}{4}( 10\,i^3 -15\,i^2 +18\,i - b),   
\end{equation}
 with $b=4$ and $b=9$ for even and odd $i$, i.e. $N_i = 1, 13, 45, 117, 239, 431,...$ (see Supplementary Note 1.2).

The path of Fig. 4a is used to design multi-species alkali clusters. The values of size mismatch between alkali atoms, estimated from nearest-neighbour distances in bulk crystals, are in the range of the optimal values for this path (Figure 4b). In addition, these species present a weak tendency to mix and the bigger atoms have a smaller cohesive energy; this produces a general tendency for the bigger atoms to stay in the surface layers, which, as we have seen in the previous section, is exactly what is needed for stabilizing icosahedral aggregates.

The clusters in Figure 4c correspond to $i=3$ (Na$_{13}$@K$_{32}$, Na$_{13}@$Rb$_{32}$) and to $i=4$ (Na$_{13}@$K$_{32}@$K$_{72}$, Na$_{13}@$Rb$_{32}@$Rb$_{72}$, Na$_{13}@$K$_{32}@$Rb$_{72}$). In these clusters, the atomic species is always changed in class-changing steps, where a large mismatch is required, while it is changed or not in class-conserving steps, where the optimal mismatch is moderate, and therefore zero mismatch is expected to be acceptable.

The energetic stability of these clusters is verified in two ways: by full global optimization searches, using a semi-empirical force field, and by density functional theory (DFT) calculations on atomic pair exchanges (see Methods for details). Our global optimization searches find that the structures designed according to the path in Figure 4 are the lowest in energy for various binary and ternary alkali systems. Complete discussion and data are reported in Supplementary Note 4.1. The DFT data are reported in Figure 5 and Supplementary Tables 1-3 of Supplementary Note 2. In the case of CK shell sequences, exchanges of pairs of different atoms in adjacent shells produce energy increases. In contrast, if MC shells of these species are assembled in the same order, the resulting clusters are energetically unstable with respect to exchanges of atomic pairs, because the mismatch between species is too large for MC shells.

We have also checked the thermodynamic stability of MC@BG and MC@BG@Ch1 structures of Figure 4c, by performing molecular dynamics simulations in which the temperature is progressively increased up to melting. Our simulations reveal that the global minimum structures are highly stable, as they do not undergo any structural transformation until the cluster entirely melts (see Supplementary Note 4.2).

Our theory therefore provides the possibility of constructing icosahedra in systems where the traditional Mackay icosahedra would not be stable. 

\subsection*{Natural growth sequences}
We have demonstrated how to geometrically construct multi-shell icosahedra and verified their energetic stability in various systems. Another important point is to understand how they grow dynamically in  physical processes. To this end, we have performed molecular dynamics (MD) growth simulations \cite{Xia2021ncomms}, in which atoms are deposited on pre-formed clusters. This type of simulations has been used to interpret previous nanoparticle growth experiments \cite{Nelli2023nanoscale,Xia2021ncomms}. The results are shown in Figure 6 and in Supplementary Figures 7-11 of Supplementary Note 3.1.

As a first case, we consider the growth of multi-element alkali clusters. In the previous section we have demonstrated the stability of non-trivial multi-shell structures, which comprise BG and chiral shells in the same aggregate. Here we check whether these structures can be grown in a physically realistic process. Here and in the following, growth temperatures are chosen to be significantly lower than the melting temperatures of the clusters.

When starting from a Na$_{13}@$Rb$_{32}$ cluster (Figure 6a), consisting of shells $(0,0)$, $(0,1)$ and $(1,1)$, our rules of 
Figure 2b predict that the next shell should be of Ch1 type, i.e. $(1,2)$ (or equivalently $(2,1)$), regardless of the size mismatch. Therefore, even when depositing atoms of the same type, we expect such symmetry breaking to take place. This is indeed the case, as shown in Figure 6a, in which in the first step of the growth a $(1,2)$ shell spontaneously form, leading to the Na$_{13}@$Rb$_{32}@$Rb$_{72}$ cluster. The symmetry breaking upon deposition of the same species of atoms is specific to BG shells. In contrast, the growth on the other type of achiral shell, i.e. MC shell, continues without symmetry breaking if atoms of the same species are deposited \cite{Nelli2023nanoscale}.

At this point, according to our rules, two non-equivalent steps are possible: the class-conserving step to $(1,3)$ and the class-changing step to $(2,2)$. These steps correspond to different optimal mismatches, i.e. 0.06 and 0.11, respectively, as estimated by Eq. (3). Since we are keeping on depositing Rb atoms (i.e. with zero mismatch), we expect growth to proceed by the $(1,3)$ step, which has the lower optimal mismatch. Also in this case, our prediction is verified by the growth simulation of Figure 6a, which further continues within the Ch1 class.   

To change class after the shell $(1,2)$, it is necessary to deposit atoms of bigger size than Rb. Specifically, the mismatch should be large enough to make the formation of the Ch1 $(1,3)$ shell unfavourable, thus addressing the growth towards the Ch2 class. At least, the mismatch between Rb and the deposited species should be larger than 0.06, which the optimal value for the formation of the Ch1 shell. From Figure 4b, it appears that Cs atoms have the right size (the mismatch between Rb and Cs is 0.08) and in fact, depositing Cs atoms on Na$_{13}@$Rb$_{32}@$Rb$_{72}$ (Figure 6b) results in a transition to the Ch2 class, as shells $(2,2)$ and then $(2,3)$ form spontaneously during the growth. 

Further confirmation of path rule predictivity is provided in Figure 6c-e. In Figure 6c, the size mismatch for the pairs AuNi, AuCo, AuFe, AgNi, AgCo, AgCu  is compared with the optimal estimates obtained from Eq. (3) for the addition of one MC and one Ch1 shell on a Mackay core. The mismatch much better corresponds to Ch1 than to MC shells. For  AuCo, AuFe, AuNi, AgNi and AgCo, Ch1 shells should grow on Mackay cores of 147 or 309 atoms ($k=3$ or $4$), while for AgCu the core should be larger, of 561 atoms ($k=5$). This is confirmed by MD simulations (Figure 6d-e) in which a Ch1 shell spontaneously forms when depositing Ag atoms on Ni$_{147}$ or Cu$_{561}$ Mackay cores. The growth continues within the Ch1 class if further Ag atoms are deposited. More results are presented and discussed in Supplementary Note 3. 

We have verified that these structures are energetically stable. Icosahedra made of a Ni, Cu or Co Mackay core surrounded by one Ch1 shell of Ag atoms are the lowest in energy in several cases \cite{Bochicchio2010nl}. When adding a second Ch1 shell, the structures are not the lowest in energy any more, but they are in close competition with the global minima (see Supplementary Note 4.1). Anyway, when heated up by molecular dynamics simulations, these MC@Ch1@Ch1 clusters preserve their structure up to the melting temperature. Data are reported in Supplementary Note 4.2.

In summary, the growth on top of icosahedral seeds naturally proceeds according to our rules for drawing paths in the hexagonal plane. At each stage of the growth, two possible steps (class-changing or class-conserving step) are possible; among them, the system spontaneously takes the step that better fits the size mismatch between atoms of the pre-existing shell and those of the growing one. If atoms of the same species are deposited, i.e. with zero mismatch, the step associated to the lower optimal mismatch is taken, which is always the class-conserving step. If one continues to deposit atoms of the same type, further and further shells belonging to the same chirality class are formed. On the other hand, the class-changing step always requires the deposition of atoms of a different species, with larger radius and with size mismatch close enough to the optimal one.

We note that almost perfect shell-by-shell growth is achieved in all simulations due to the fast diffusion of deposited atoms on top of the close-packed shells \cite{Nelli2023nanoscale}.

Here we have shown and discussed the growth at some selected simulation temperatures. In Supplementary Note 3.4 we discuss the effect of temperature on the growth, showing that this type of icosahedral growth process is very robust, as it takes place in a wide range of temperatures.

\subsection*{Extension to anti-Mackay shells}
The mapping of icosahedral structures into paths can be extended to other cases. Here we establish the extension to the generalized anti-Mackay (AM) icosahedral shells of Figure 7. 
AM shells are achiral and not close-packed, since they contain non-vertex particles with coordination lower than six. Each AM shell is identified by a pair of non-negative integers, which here we call $(p,q)$, determining the disposition of particles in the triangular facet of the shell (see Figure 7a). Shells with $q=0$ have been described in the original work of Mackay \cite{Mackay1962ac}, who proposed the possibility of adding to the fcc tetrahedra of the Mackay icosahedron one more shell in hexagonal close-packed (hcp) stacking. Multi-shell icosahedra terminated by AM shells of different types have been observed in confined aggregates of colloidal particles \cite{deNijs2015natphys,Chen2021natphys,Wang2019acsnano}.

In Supplementary Note 1.5, we demonstrate that there is a correspondence between AM and CK shells (of chiral and BG type), as an AM shell of indexes $(p,q)$ has the same number of particles of the CK shells with $h=q+1$, $k=p+q$, if $k>h$, and $k=q+1$, $h=p+q$ if $h>k$ ($p=1$ gives the BG shells, that are common to AM and CK structures). Therefore we can unambiguously identify an AM shell by the $(h,k)$ indexes of a corresponding CK shell, and assign to it the same lattice point on the hexagonal plane. We group AM shells into AM$n$ classes and denote an AM shell by $(h,k)^*$. The correspondence between AM$n$ and Ch$n$ shells is explicitly shown in Figure 7b,c for $n=1,2$.

AM shells can be packed by using the same rules described for CK ones. The same elementary steps in the hexagonal plane are allowed, but now one can decide whether to consider the CK or the AM shell corresponding to the endpoint of the step. In this way a larger variety of icosahedra can be built. Again, the stability of these structures is ruled by the size mismatch between particles in different shells. 
The optimal size mismatch for icosahedra with AM shells can be estimated by using the same type of geometric considerations made for CK shells (see Supplemmentary Note 1.5).

In Figure 8a we compare the stability of all possible shells that can be put on top of a Mackay core, namely MC, Ch1 and AM1. The optimal mismatch of the AM1 shells is intermediate between those of the MC and Ch1 shells (see also Supplementary Figure 3). The mismatch for adding a AM1 shell on a 147-atom Mackay core is close to the one of AgCu, so that an AM shell should grow by depositing Ag atoms on Cu a Mackay core of this size. This is verified by the MD simulations of Figure 8b, in which we observe the formation of an Ag $(1,3)^\ast$ shell.

From there on, the growth proceeds in a different way from that of Figure 6e: a second shell of AM2 type is formed, followed by a shell of BG type, i.e. the class is changed at each step, even though atoms of the same type are deposited. This behaviour is due to the kinetic of the growth process. Specifically, it is due to fourfold adsorption sites on the surface of AM shells, which act as adatom traps and naturally lead to the formation of AM shells of higher class. The fourfold traps hinder the atomic mobility, causing a much rougher growth than in the case of Figure 6e. This point is discussed in detail in Supplementary Note 3.3. We note that, even though kinetic effects dominate the process, the growth proceeds step by step by incrementing only one shell index, as predicted by our rules.

Finally, in Figure 8c we show the growth of a ternary NiPdAg cluster. We start by depositing Pd atoms on a Ni$_{147}$ Mackay cluster, that spontaneously arrange into an anti-Mackay $(1,3)^\ast$ shell. The mismatch between Ni and Pd is of 0.10, i.e. it is much larger than the optimal mismatch for growing a MC shell (0.04), and quite close to the optimal mismatch for a AM one (0.12, as estimated by geometrical considerations). On the other hand, the optimal mismatch for the formation of a Ch1 shell is quite larger (0.14, see Figure 8a). If the growth is continued by depositing Ag atoms, shells $(2,3)^\ast$ and then $(3,3)$ form. Such AM growth pathway appears to be unaffected by the mismatch, since it is observed also in the case of zero mismatch of Figure 8b. However, as for the CK shells, one can estimate the optimal size mismatch for changing the AM class, finding values sensibly larger than zero, and in the same range of those calculated for the CK shells. Therefore, depositing atoms of larger size, such as Ag (the mismatch between Pd and Ag is of 0.05), is expected to be beneficial for the stability of the resulting aggregate. Further deposition of Ag atoms leads to the formation of a chiral Ch3 shell, of type $(3,4)$. Indeed, the $(3,3)$ BG shell can be seen both as a generalised AM and a CK shell, and therefore it allows to obtain a very peculiar growth sequence, in which the two families of icosahedral shells are present in the same structure. We remark that the growth on top of a BG shell naturally leads to the formation of a chiral shell instead of an AM one, due to the lack of fourfold adsorption sites that are needed to form AM shells of class larger than 1.

From the energetic point of view, also the generalized anti-Mackay structures may present notable stability. In fact, Ni$_{147}$@Pd$_{132}$ and Ni$_{147}$@Pd$_{132}$@Ag$_{192}$ generalized anti-Mackay structures are the lowest in energy (see also Supplementary Note 4.1).

\section*{Discussion}
Here we discuss the potential applications of the icosahedral structures described in this work, and we suggest possible extensions of our design strategy.

In many cases, the applications of nanoparticles stem from the structure of their surface. Our work shows that, to obtain a specific type of surface shell, it is necessary to build the correct sequence of shells in the inner part of the nanoparticles. This is possible if different elements (two or more) are used, with the appropriate mismatch. Applications of different kinds are also possible. For metal nanoparticles, it may be useful to protect a core (for example a magnetic core made of Ni, Co or Fe) by covering it with another metal. Our results indicate that for this purpose the size mismatch must be carefully chosen, depending also on the size of the magnetic core. If not, other asymmetric quasi-Janus structures with off-centered core would be much more stable (for example see ref. \cite{Bochicchio2013prb} and Supplementary Figure 15), and the core would be protected much less effectively.

In this work, we have treated specific examples related to metal clusters and nanoparticles. However, we believe that our construction can be relevant for other systems, e.g. colloidal aggregates. Colloids interact by potentials with very short range \cite{Cerbelaud2024mtc}. In Supplementary Note 1.6 we have shown that, for systems with short-range interaction, it is still possible to build stable structures according to our rules. However, the structures are less tolerant of deviations from the optimal size mismatch between shells, therefore being stable in a narrower range of mismatch values around the optimal one. This implies that shells must be carefully assembled to satisfy the constraints on mismatch values. In addition, we note that in some experiments on hard-sphere colloids, in which particle-particle attractive forces are negligible, icosahedral structures have been obtained in confined environments due to entropic effects \cite{deNijs2015natphys,Wang2018ncomms,Chen2021natphys}. These experiments were considering colloids of equal sizes and obtained aggregates with Mackay or anti-Mackay arrangements. Our results may be relevant also for these systems, since they may give a systematic guide on how to build those confined colloidal systems using colloids with different sizes. In fact, our approach, which is mostly based on general geometric considerations, is able to indicate the appropriate size mismatch between the spheres.

A natural generalization of the results presented here concerns in drawing paths that do not obey to the step rule $(h,k) \to (h,k+1)$ or $(h,k) \to (h+1,k)$. These paths are discussed in Supplementary Note 5, where we show that they are less likely to be physically relevant because they lead to poor matching between atoms in adjacent shells.

Our construction can be further generalized. Let us mention a few possibilities. First, the paths can begin at any point in the lattice, instead of (0,0), so that the shells enclose an empty volume. Multi-shell structures enclosing a cavity are relevant to biological systems \cite{Zandi2020pr,Prinsen2010jpcb,Sasaki2016jpcb}, and have been observed in metal clusters \cite{Whetten2019acr}.
Second, the non-equivalent sites of a shell can be decorated with different types of particles, while maintaining icosahedral symmetry. For example, the vertex atoms in the surface shell of a metal cluster can be of a different species than the other atoms, thus becoming isolated impurities embedded in a surface of a different material. This possibility is relevant to single-atom catalysis \cite{Wang2018nrc}. 
Moreover, the mapping into paths may be applied to other figures obtained by cutting and folding the hexagonal lattice, such as octahedra and tetrahedra, and to other lattices, including Archimedean lattices \cite{Twarock2019ncomms}, that can better accommodate particles with non-spherically symmetric interactions. 

In summary, the mapping into paths is a powerful tool for the bottom-up design of chiral and achiral aggregates of atoms, colloids and complex molecules. The unusual geometries of these aggregates can be of interest in various fields, e.g. in catalysis, optics and synthetic biology.

\section*{Methods}
\subsection*{Construction of the structures}
Icosahedral structures are built by assembling Caspar-Klug and anti-Mackay shells by our C++ code. The code takes as input the geometric features of each shell, i.e. the indexes $h$ and $k$ of the CK construction, the distance $d$ between particles, and the shell type (either CK or AM). For CK shells, the first 12 particles are placed in the vertices of the icosahedron of edge legth $\sin(2\textgreek{π} /5)\sqrt{T}\, d$. The other particles are placed on the icosahedral facets according to the CK scheme: on each facet plane a 2D hexagonal lattice is built, which is rotated of an angle of amplitude $\theta_\text{CK}$ with respect to one of the facet edges; particles are placed on lattice points falling within the facet. For AM shells, the indexes $p$ and $q$ are calculated as $p=k-h+1$, $q=h-1$ if $k\geq h$, $p=h-k+1$, $q=k-1$ otherwise. The first 12 particles are placed in the vertices of the icosahedron of edge length $(p+\sqrt{3}\,q + \sqrt{3}-1)\,d$. The other particles are placed on the icosahedral facets according to scheme of Figure 6a. A more detailed description of CK and AM shells and of the procedure for constructing them can be found in Supplementary Note 1.

\subsection*{LJ and Morse potential calculations} 
Both potentials are pair potentials, in which the total energy $E$ is written as
\begin{equation}
    E= \frac{1}{2} \sum_{i\neq j} u(r_{ij}),
\end{equation}
where $r_{ij} =|\mathbf{r}_j-\mathbf{r}_j|$ is the distance between a pair of particles.
The LJ potential is written as
\begin{equation}
 u(r) =  \varepsilon \left[ \left( \frac{r_\text{m}}{r}\right)^{12} - 2 \left( \frac{r_\text{m}}{r}\right)^{6}\right], 
\end{equation}
 where $\varepsilon$ is the well depth and $r_\text{m}$ is the equilibrium distance, which corresponds to the particle size $d$.
The Morse potential is written as
\begin{equation}
 u(r) =  \varepsilon \left[ \mathrm{e}^{-2\alpha (r/r_\text{m}- 1)} -2 \mathrm{e}^{-\alpha (r/r_\text{m}-1)}\right], 
\end{equation}
where the dimensionless parameter $\alpha$ regulates the width of the potential well, that decreases with increasing $\alpha$. For $\alpha=6$ both LJ and Morse potential have the same width of the well, i.e. the same curvature at the well bottom. For $\alpha < 6$ the potential well is wider than in LJ. The chosen values $\alpha=4,5$ give widths of the potential well similar to those for the interaction between metal atoms.
In the simulations of Figures 3a,c,d and 8a all particles were given the same value of $\varepsilon$ (and of $\alpha $ for the Morse potential), but particles of the outer shell and of the core were given different sizes (i.e. $r_\text{m}=r_\text{m,c}$ for core particles and $r_\text{m}=r_\text{m,s}$ for outer shell particles). For interactions between particles of the core and the outer shell $r_\text{m,cs}=(r_\text{m,c}+r_\text{m,s})/2$. The structures were locally relaxed by quenched molecular dynamics \cite{Bennett1975} to reach the position of the local minimum in the energy landscape.

\subsection*{Global optimization searches}
Global optimization searches are performed by the Basin Hopping algorithm \cite{Wales1997jpca} and its extensions \cite{Rossi2009jpcm,Rapetti2023ats}. In all cases, at least four independent unseeded simulations of 1-4$\times$10$^6$ steps were performed, plus some seeded simulations starting from selected structures. All global minima reported here and in Supplementary Note 4.1 resulted from unseeded simulations. For all systems, atom-atom interactions were modelled by an atomistic force field, which is known as Gupta potential \cite{Gupta1981prb}.
Form and parameters of the potential can be found in refs. \cite{Li1998prb,Baletto2002prb,Baletto2003prl,Rossi2009jctn}.

\subsection*{DFT calculations}
All DFT calculations were made by the open-source QUANTUM ESPRESSO software \cite{QE-giannozzi} using
the Perdew-Burke-Ernzerhof exchange-correlation functional \cite{Perdew1996prl}. The convergence thresholds
for the total energy, total force, and for electronic calculations were set to 10$^{-4}$ Ry, 10$^{-3}$ Ry$\,$at.u$^{-1}$. and 5 $\times$ 10$^{-6}$ Ry respectively. We used a periodic cubic cell, whose size was set to 26-48 \AA, depending on the size of the cluster, in order to ensure at least a 10 \AA \, separation between clusters in different periodic images. Cutoffs for wavefunction and charge density were set to 66 and 323 Ry, according
to Na.pbe-spn-kjpaw\_psl.1.0.0.UPF, K.pbe-spn-kjpaw\_psl.1.0.0.UPF, Rb.pbe-spn-kjpaw\_psl.1.0.0.UPF as provided by the QUANTUM ESPRESSO pseudopotential library available at \\
\url{http://pseudopotentials.quantum-espresso.org/legacy\_tables/ps-library/}.

\subsection*{MD growth simulations}
Molecular dynamics growth simulations are made by molecular dynamics using the same type of procedure adopted in refs. \cite{Nelli2023nanoscale,Xia2021ncomms}. The equations of
motion are solved by the Velocity Verlet algorithm with a time step of 5 fs for the simulations of AgNi, AgCu, AgCo, AuCo, AuFe, AgPdNi and 2 fs for the simulations of NaK, NaRb, NaKRb, NaRbCs. In all simulations, the temperature is kept constant by an Andersen thermostat with a collision frequency of 5$\times$10$^{11}$ s$^{-1}$. Simulations start from a seed, which is an initial cluster, then atoms are deposited one by one on top of it in an isotropic way from random directions at a constant rate. The simulation of Figure 6a in the main text was started from a Na$_{13}@$K$_{32}$ Bergman-type seed corresponding to the path arriving to $(h,k)=(1,1)$ and Rb atoms were deposited at a rate of 0.1 atoms$\,$ns$^{-1}$ and at a temperature of 125 K. 
The simulation of Figure 6b text was started from a Na$_{13}@$Rb$_{32}@$Rb$_{72}$ chiral seed corresponding to the path arriving to $(h,k)=(1,2)$ and Cs atoms were deposited at a rate of 0.1 atoms$\,$ns$^{-1}$ at 125 K. 
The simulation of Figure 6c was started from a Ni$_{147}$ Mackay icosahedral seed corresponding to the path arriving to $(h,k)=(0,3)$ and Ag atoms were deposited at a rate of 0.1 atoms$\,$ns$^{-1}$ at 450 K. 
The simulation of Figure 6d was started from a Cu$_{561}$ Mackay icosahedral seed corresponding to the path arriving to $(h,k)=(0,5)$ and Ag atoms were deposited at a rate of 0.1 atoms$\,$ns$^{-1}$ at 450 K.
The simulation of Figure 8b was started from a Cu$_{147}$ Mackay icosahedral seed corresponding to the path arriving to $(h,k)=(0,3)$ and Ag atoms were deposited at a rate of 1 atoms$\,$ns$^{-1}$ at 350 K.
The simulation of Figure 8c were started from a Ni$_{147}$ Mackay icosahedral seed corresponding to the path arriving to $(h,k)=(0,3)$ and Pd atoms were deposited at a rate of 0.1 atoms$\,$ns$^{-1}$ at 400 K. Then Ag atoms were deposited on a Ni$_{147}$@Pd$_{132}$ seed terminated by a $(1,3)^\ast$ AM1 shell, at a rate of 0.1 atoms$\,$ns$^{-1}$ at 300 K.
For all systems, atom-atom interactions were modelled by an atomistic force field, which is known as Gupta potential \cite{Gupta1981prb}.
Form and parameters of the potential can be found in refs. \cite{Li1998prb,Baletto2002prb,Baletto2003prl,Rossi2009jctn}.

\section*{Data Availability}
The data that support the findings of this study are available from the corresponding authors upon request. Source data are provided as a Source Data file.

\section*{Code Availability}
Codes used in this study, such as for the MD simulations and for constructing the multi-shell icosahedra, are available from the corresponding authors upon request.




\section*{Acknowledgments}
The authors acknowledge financial support under the National Recovery and Resilience Plan (NRRP), Mission 4, Component 2, Investment 1.1, Call for tender No. 104527 published on 2.2.2022 by the Italian Ministry of University and Research (MUR), funded by the European Union – NextGenerationEU– Project Title PINENUT – CUP D53D23002340006 - Grant Assignment Decree No. 957 adopted on 30/06/2023 by the Italian Ministry of University and Research (MUR). The authors thank Giovanni Barcaro, El yakout El koraychy, Alessandro Fortunelli, Alberto Giacomello, Laurence D. Marks, Mauro Moglianetti, Richard E. Palmer, Emanuele Panizon, Michele Parrinello, Cesare Roncaglia, Giulia Rossi, Manoj Settem, Erio Tosatti and Jeffrey R. Weeks for a critical reading of the manuscript.

This version of the article has been accepted for publication, after peer review but is not the Version of Record and does not reflect post-acceptance improvements, or any corrections. The Version of Record is available online at: \\
\url{https://doi.org/10.1038/s41467-025-56952-1}

\section*{Author Contributions Statement}
NC performed and analyzed the growth simulations. DN developed the analytical calculations of mismatch, made the DFT calculations and developed the code for constructing chiral and anti-Mackay shells. RF made Lennard-Jones and Morse calculations. DN and RF made the global optimization searches and supervised the work. All authors contributed to writing the paper.

\section*{Competing Interests Statement}
The authors declare no competing interests.

\section*{Figures}

\begin{figure}
    \centering
    \includegraphics[width=\textwidth]{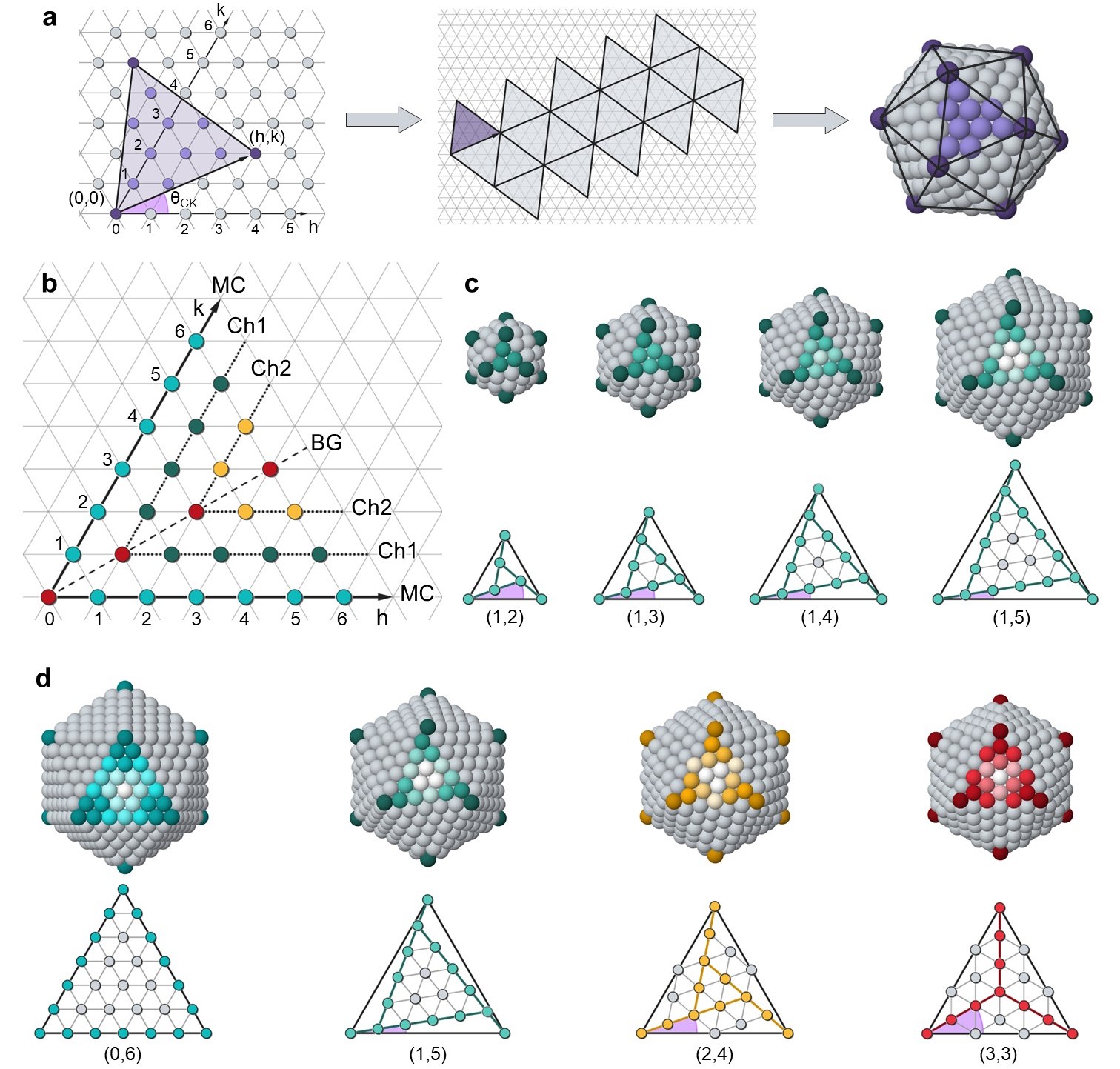}
    \caption{}
\end{figure}

\begin{figure}
    \centering
    \includegraphics[width=\textwidth]{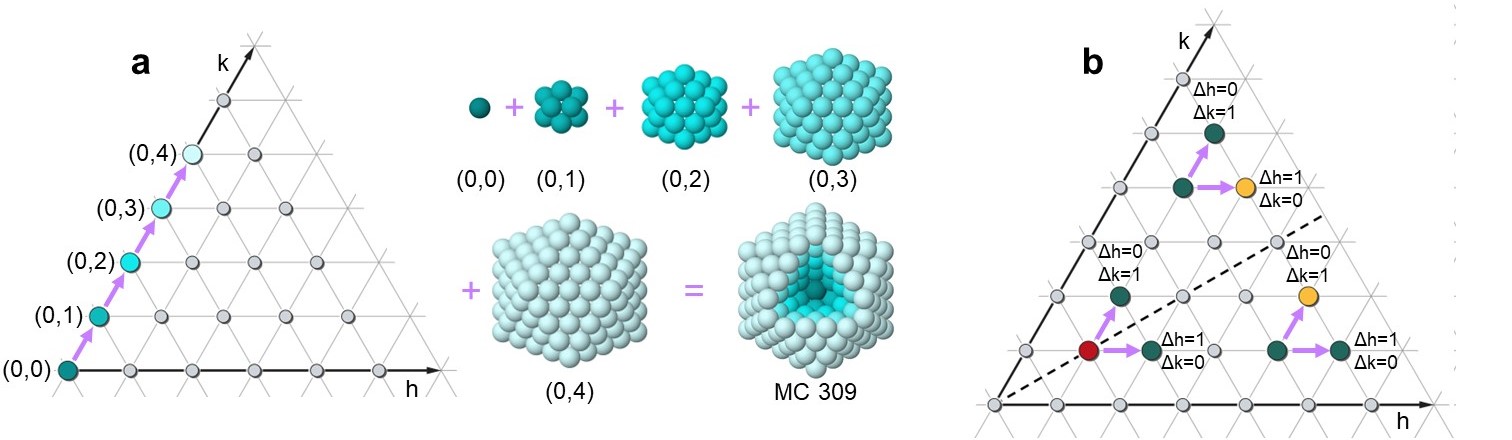}
    \caption{}
\end{figure}

\begin{figure}
    \centering
    \includegraphics[width=\textwidth]{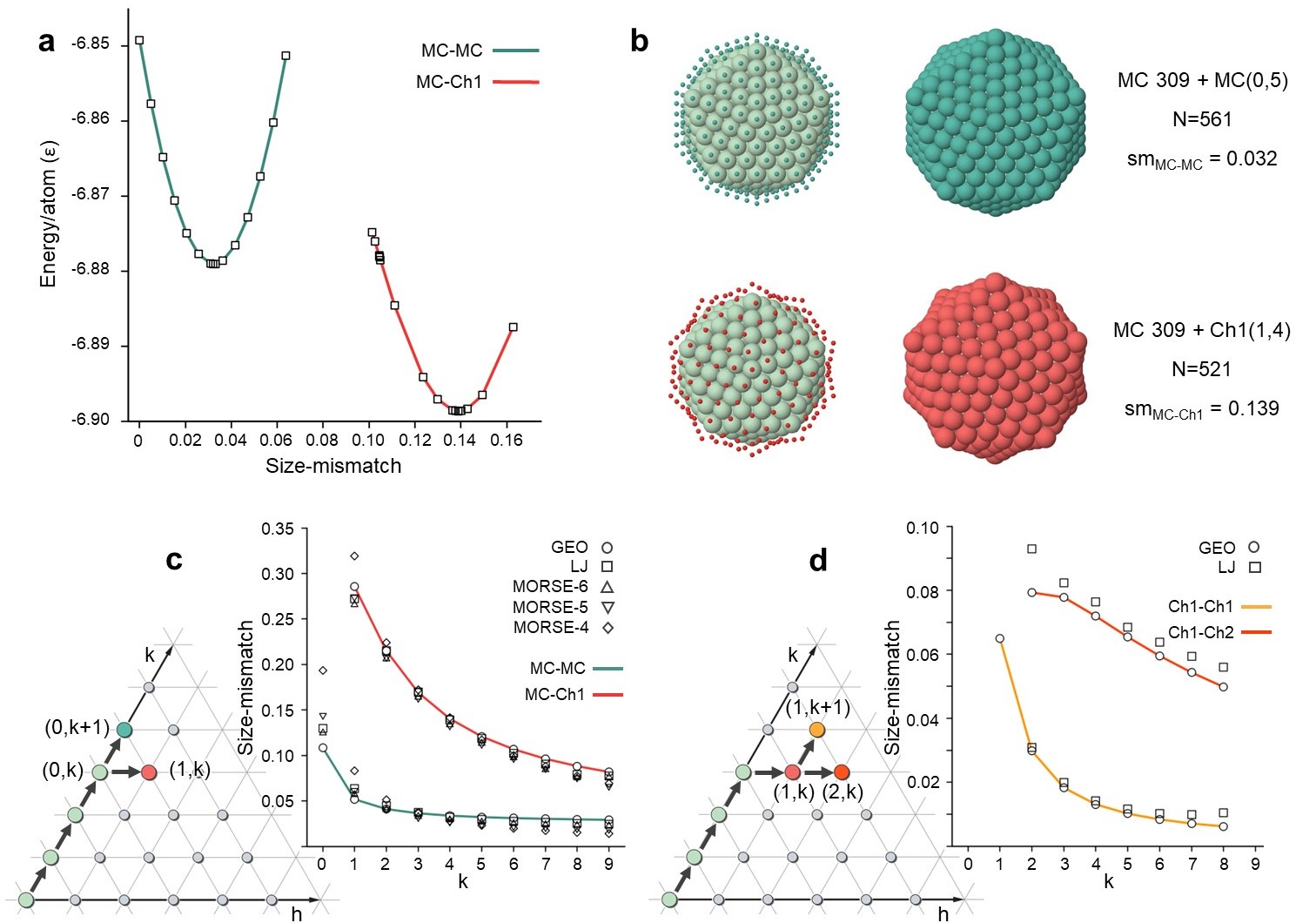}
    \caption{}
\end{figure}

\begin{figure}
    \centering
    \includegraphics[width=\textwidth]{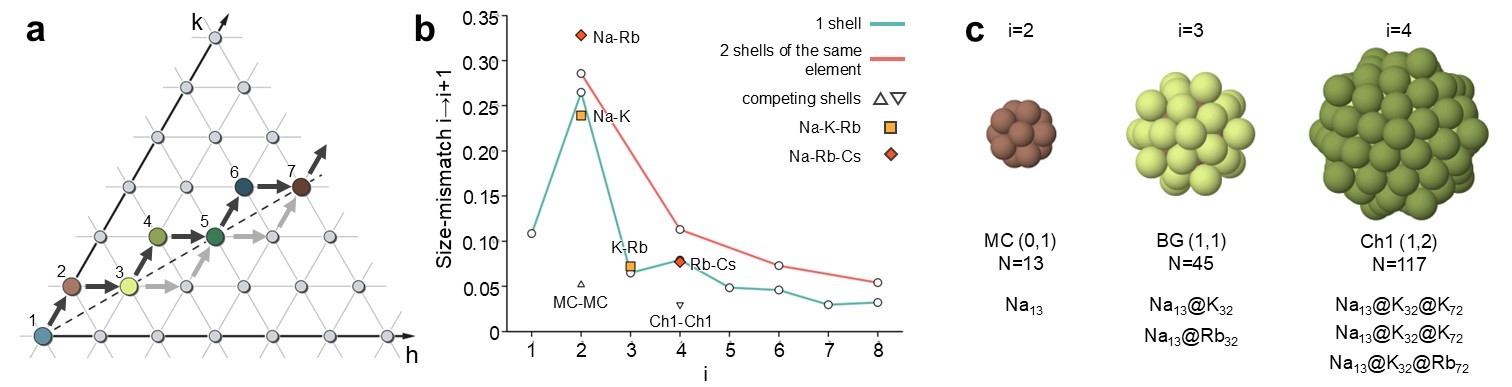}
    \caption{}
\end{figure}

\begin{figure}
    \centering
    \includegraphics[width=\textwidth]{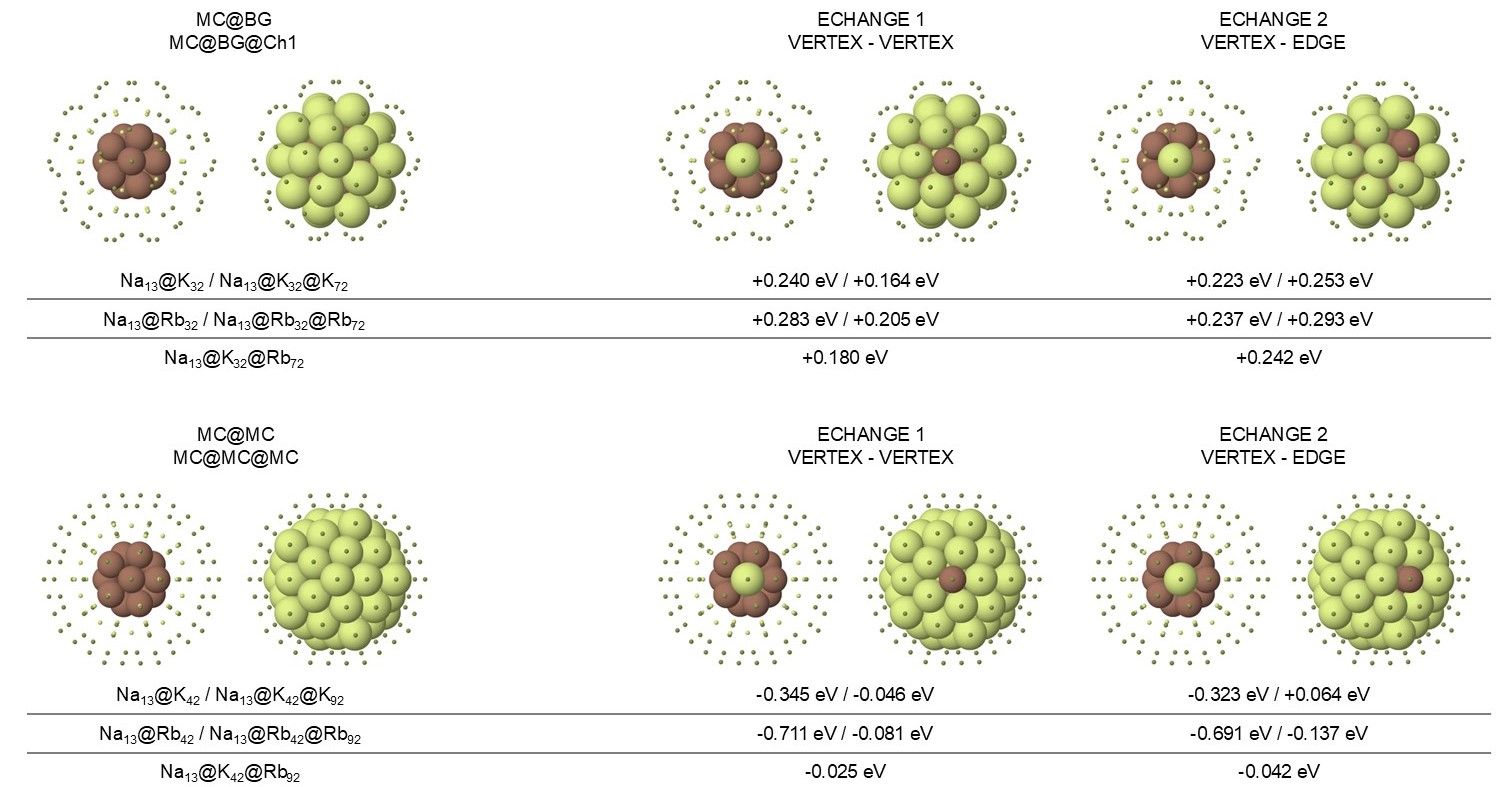}
    \caption{}
\end{figure}

\begin{figure}
    \centering
    \includegraphics[width=\textwidth]{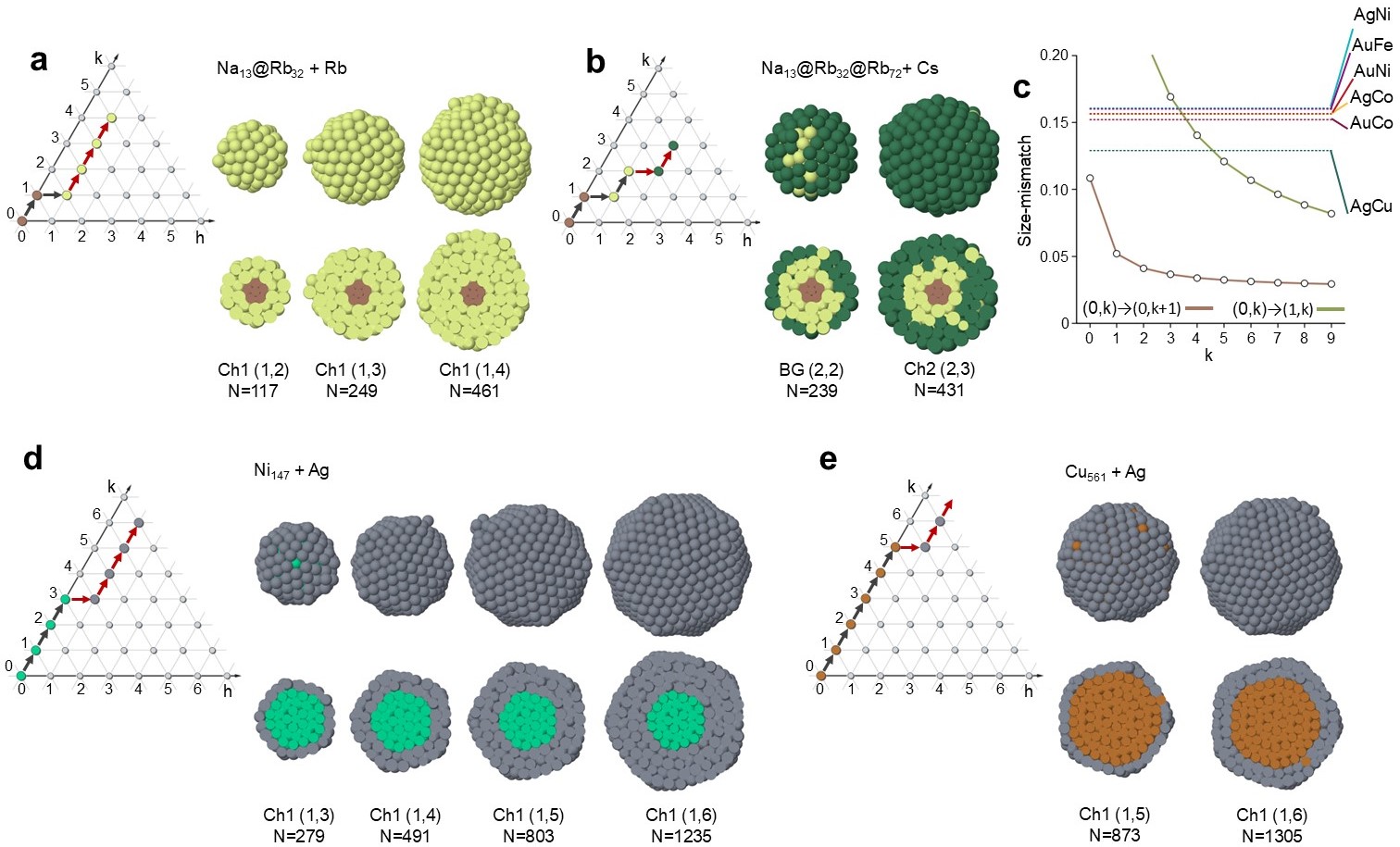}
    \caption{}
\end{figure}

\begin{figure}
    \centering
    \includegraphics[width=\textwidth]{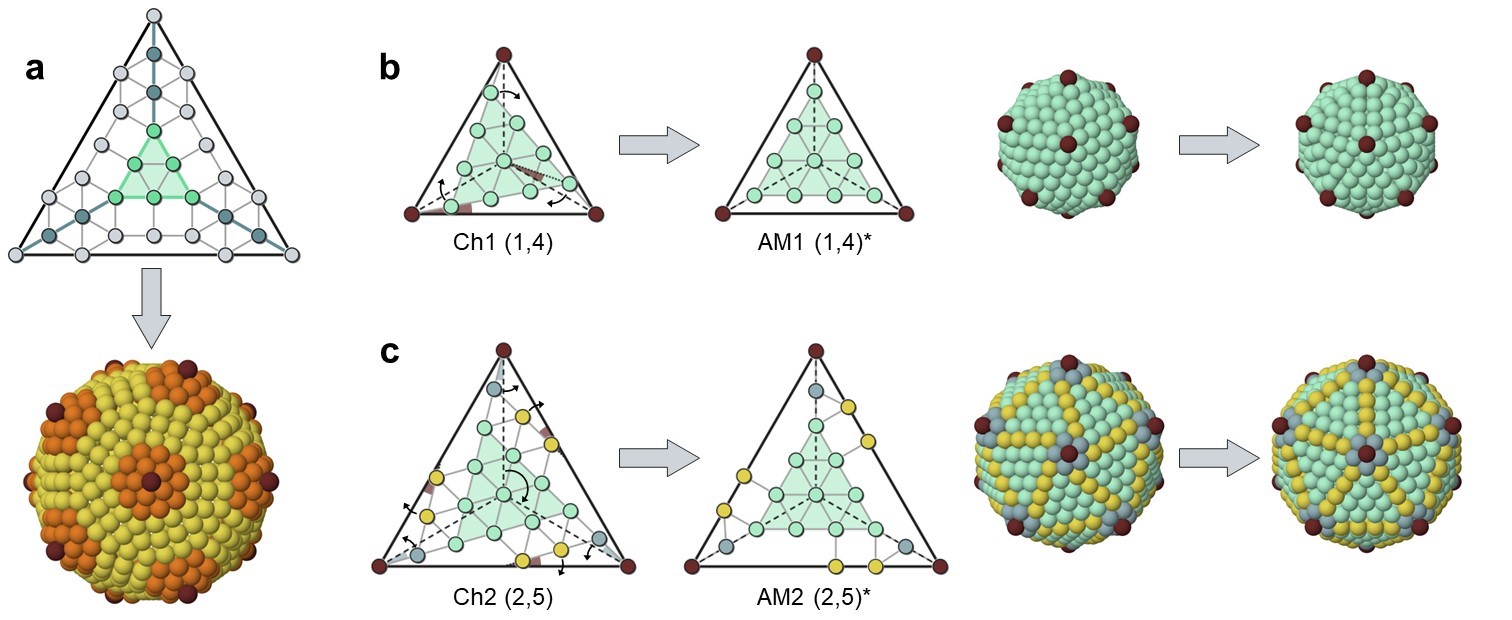}
    \caption{}
\end{figure}

\begin{figure}
    \centering
    \includegraphics[width=\textwidth]{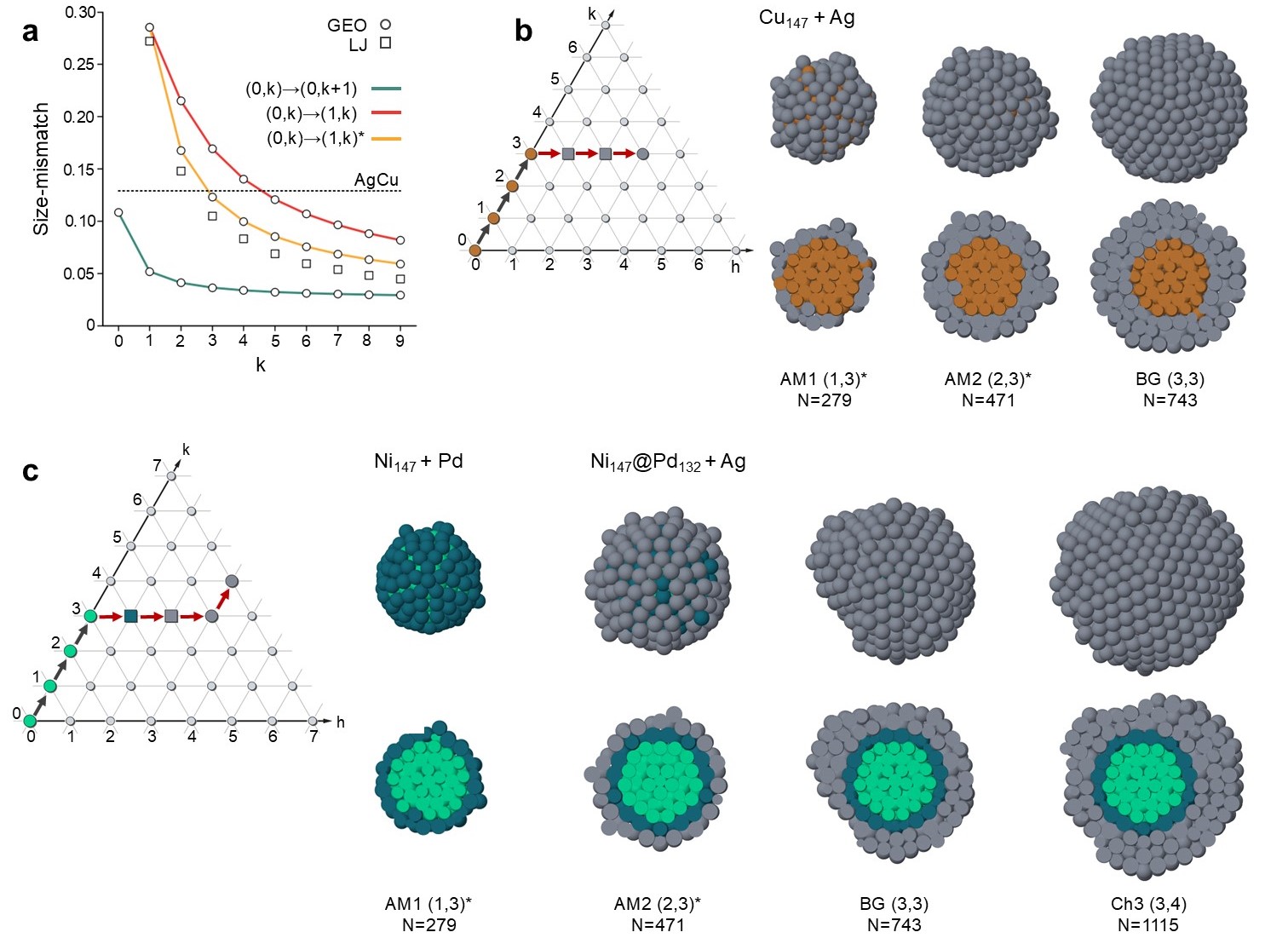}
    \caption{}
\end{figure}

\section*{Figure Legends/Captions}

\subsection*{Figure 1}
\textbf{Caspar-Klug shells and chirality classes.} \textbf{a} Caspar-Klug construction \cite{Caspar1962cshs,SadreMarandi2018cmb,Twarock2019ncomms}. The coordinate axes $h$ and $k$ ($(h,k)$ non-negative integers) are at 60$^\circ$. From left to right: a segment from $(0,0)$ to $(h,k)$ is drawn and an equilateral triangle is constructed on it; the triangle is repeated 20 times to form a leaflet that is cut and folded into an icosahedral shell. \textbf{b} Correspondence between $(h,k)$ points and icosahedral shells. The points on the coordinate axis (light blue) correspond to achiral MC shells \cite{Mackay1962ac,Baletto2005rmp}, those on the diagonal (red) to achiral BG shells \cite{Bergman1957ac,Pankova2015ic}. All other points correspond to chiral shells. Shells are grouped into classes Ch$n$ as explained in the text (MC$\equiv$Ch0). Green and red points correspond to Ch1 and Ch2 shells, respectively. \textbf{c} The first four right-handed shells of the Ch1 class. The triangles in the bottom row (identified by their $(h,k)$) show the angle $\theta=60^\circ-\theta_\text{CK}$ between the facet edge and the line connecting the vertex to a nearest neighbour point, that decreases with increasing $k$. \textbf{d} Achiral and right-handed shells with $h+k=6$. From left to right, MC, Ch1, Ch2 and BG shells. $\theta$ increases from 0$^\circ$ to 30$^\circ$ from MC to BG. In the top rows of \textbf{c} and \textbf{d} the color shades identify symmetrically equivalent particles.

\section*{Figure 2}
\textbf{Mapping icosahedra into paths.} \textbf{a} Path of a Mackay icosahedron: $i=5$ shells are assembled on top of each other. \textbf{b} The three possible cases when choosing between $(h,k) \to (h,k+1)$ and $(h,k) \to (h+1,k)$ at each step of the path. In \textbf{b}, points are colored according to the chirality class of the corresponding shell, as in Figure 1b.

\section*{Figure 3}
\textbf{Optimal size mismatch.} \textbf{a} Binding energy per atom for clusters made of a core with five MC shells plus a sixth shell, which is either MC (blue curve) or Ch1 (red curve). Interactions are of Lennard-Jones (LJ) type (see Methods). The particles of the sixth shell differ from those in the core only by their size. The energy is in units of the $\varepsilon$ of the LJ potential. The structures are shown in \textbf{b}. \textbf{c} Comparison of the optimal mismatch between steps $(0,k) \to (0,k+1)$ (blue curve) and $(0,k) \to (1,k)$ (red curve), corresponding to additional MC and Ch1 shells on an MC core, respectively. The optimal mismatch of Eq. (3) (GEO values) is compared to LJ and Morse potential results. The Morse potential data are given for three values (4,5,6) of the parameter $\alpha$, that regulates the width of the potential well (see the Methods section). \textbf{d} Comparison between steps $(1,k) \to (1,k+1)$ (yellow curve) and $(1,k) \to (2,k)$ (orange curve), corresponding to additional Ch1 and Ch2 shells on a core containing $k$ MC and one Ch1 shells, respectively. The optimal mismatch of Eq. (3) (GEO) is compared to LJ data. Source data of \textbf{a}, \textbf{c}, \textbf{d} are provided as a Source Data file.

\section*{Figure 4}
\textbf{A path with spontaneous symmetry breaking.} \textbf{a} A path connecting all BG shells through their nearby chiral shells. Black and grey arrows indicate paths with right and left-handed chiral shells. A generic path of this type may alternate both chiralities, corresponding to combinations of black and grey arrows. Shells are enumerated along the black-arrow path. \textbf{b} Optimal mismatch $\text{sm}_{i,i+1}$ according to Eq. (3) for the path in \textbf{a}. Points connected by blue and red lines correspond to changes of species every shell and every two shells. The optimal values of $\text{sm}_{i,i+1}$ are compared to the mismatch between pairs of atomic species, indicated by squares and diamonds. The triangles correspond to the optimal mismatch between two MC shells and two Ch1 shells. \textbf{c} Structures for $i=2,3,4$ along the path of \textbf{a}, with their compositions related to specific systems. Source data of \textbf{b} are provided as a Source Data file.

\section*{Figure 5}
\textbf{DFT data for atomic pair exchanges.} Perfect structures before the atomic pair exchange, either of \textbf{a} MC@BG/MC@BG@Ch1 or \textbf{d} MC@MC/MC@MC@MC type. Atoms in the second and third shell are colored in brown and yellow, respectively. \textbf{b}, \textbf{c}, \textbf{e}, \textbf{f} configurations after the atomic pair exchanges of two different types. Energy differences with respect to the perfect configurations in \textbf{a} and \textbf{d} are reported below. Positive and negative energy differences indicate unfavourable and favourable exchange processes, respectively. Complete data are reported in Supplementary Tables 1-3.

\section*{Figure 6}
\textbf{Growth sequences of chiral icosahedra.} \textbf{a} Rb atoms are deposited on an icosahedral Na$_{13}@$Rb$_{32}$ seed. \textbf{b} The seed is Na$_{13}@$Rb$_{32}$@Rb$_{72}$, on which Cs atoms are deposited. \textbf{c} Size mismatch for some transition metal pairs compared to the optimal mismatch for a MC and a Ch1 shell on a MC core (brown and green lines, respectively), as a function of core size. \textbf{d-e} Snapshots from growth simulations and corresponding paths for the deposition of Ag atoms on \textbf{d}  Ni$_{147}$ and \textbf{g} Cu$_{561}$ cores. All simulation snapshots in \textbf{a-b}, \textbf{d-e} are taken at magic sizes for the corresponding paths. In the top and bottom rows of the snapshot sequences we show the cluster surface and its cross section, respectively. In the representation of shell sequences in the hexagonal plane, black arrows are used to connect the shells belonging to the initial icosahedral seed, while red arrows connect the shells spontaneously formed on top of it in the growth simulations. Na, Rb, Cs, Ni, Cu and Ag atoms are colored in brown, yellow, dark green, light green, orange and grey, respectively. Source data of \textbf{c} are provided as a Source Data file.

\section*{Figure 7}
\textbf{Generalized anti-Mackay shells.} \textbf{a} A shell of the AM family. A facet and the complete shell are shown in the upper and lower panels. The shell is identified by $(p,q)$, with $p$ number of particles on the side of the inner triangle (light green particles in the top panel) and $q$ number of particles between nearby vertices of the outer and inner triangles (blue particles). In this case, $(p,q)=(3,2)$. In the bottom panel, orange particles have coordination 6 within the shell, while other particles have lower coordination. \textbf{b} AM1$\leftrightarrow$Ch1 and \textbf{c} AM2$\leftrightarrow$Ch2 correspondences. AM1$\leftrightarrow$Ch1 amounts to the rotation of the inner triangles \cite{Bochicchio2010nl}, whereas AM2$\leftrightarrow$Ch2 involves rotations of different groups of particles, represented by different colours. A more detailed description is in Supplementary Note 1.5.

\section*{Figure 8}
\textbf{Growth of generalized anti-Mackay structures.} \textbf{a} Optimal mismatch for one AM1 shell on a Mackay core ($(0,k)\to(1,k)^*$) as a function of core size (yellow curve), compared with Ch1 ($(0,k)\to(1,k)$) (red curve) and Mackay shells ($(0,k)\to(0,k+1)$) (blue curve), for Lennard-Jones clusters. GEO values are calculated by the formula derived in Supplementary Note 1.5. The mismatch between Ag and Cu is indicated. \textbf{b},\textbf{c} Snapshots from  MD growth simulations (with cluster surfaces and cross sections shown in top and bottom lines, respectively) and corresponding paths, with AM shells indicated by squares. In \textbf{b} Ag atoms are deposited on a Mackay Cu$_{147}$ core. In \textbf{c} Pd atoms are deposited on a Ni core and then Ag atoms are deposited on the Ni@Pd cluster. In the top and bottom rows of the snapshot sequences we show the cluster surface and its cross section, respectively. In the representation of shell sequences in the hexagonal plane, black arrows are used to connect the shells belonging to the initial icosahedral seed, while red arrows connect the shells spontaneously formed on top of it in the growth simulations. Ni, Cu, Pd and Ag atoms are colored in light green, orange, blue and grey, respectively. Source data of \textbf{a} are provided as a Source Data file.

\clearpage



\section*{Supplementary material}

# Supplementary Material for Path-based packing of icosahedral shells into multi-component aggregates

Nicolò Canestrari, Diana Nelli,* Riccardo Ferrando*

Dipartimento di Fisica, Università di Genova, Via Dodecaneso 33, 16146 Genova, Italy

*To whom correspondence should be addressed;
E-mail: diana.nelli@edu.unige.it; riccardo.ferrando@unige.it.

## S1 Construction of structures and optimal mismatch

### S1.1 Number of particles in a CK shell

The number of particle in a Caspar-Klug (CK) shell can be calculated from the ratio between the area of the triangular facet of the icosahedron and the area occupied by a particle on the CK plane. Icosahedral facets are equilateral triangles of side length $\sqrt{T}$, $T$ being the triangulation number of the CK shell. Therefore the area of the facet is

$$A_T = \frac{1}{2}\sqrt{T}\sqrt{T}\sin 60° = \frac{1}{2}T\frac{\sqrt{3}}{2} = \frac{\sqrt{3}}{4}T. \tag{S1}$$

The area occupied by a particle corresponds to the area of the unit cell of the 2D hexagonal lattice with unit distance between two nearby lattice points, which is

$$A_P = \frac{\sqrt{3}}{2}. \tag{S2}$$

Therefore the number of particles in a triangular facet of the CK icosahedron is

$$N_T = \frac{A_T}{A_P} = \frac{\frac{\sqrt{3}}{4}T}{\frac{\sqrt{3}}{2}} = \frac{T}{2}. \tag{S3}$$

To calculate the total number of particles in the shell we multiply $N_T$ by the number of facets in the icosahedron, which is 20. However, we need to be careful when considering icosahedral

vertices. In the calculation of $N_T$, the number of particles assigned to each vertex is 1/6, since it is shared among 6 equilateral triangles in the hexagonal lattice, while in the icosahedron, each vertex is shared among 5 facets. As a consequence, if we calculate the total number of particles as $20 N_T$, we are assigning to each vertex 5/6 particles instead of 1. Therefore we need to add further 1/6 particles per vertex. Since the icosahedron has 12 vertices, 2 particles have to be added. The number of particles in the CK icosahedral shell is therefore

$$N_{SHELL} = 20\,N_T + 2 = 10\,T + 2. \tag{S4}$$

## S1.2  Magic numbers

Eq. S4 can be used to calculate the magic numbers of any icosahedral series, i.e. of icosahedra mapped into any pathway in the closed-packed plane. We recall that magic numbers are those sizes at which perfect icosahedra can be built, and therefore correspond to the completion of the different concentric CK shells. Here we calculate the magic numbers of two interesting icosahedral series, i.e.

1. the series in Figure 4a in the main text, passing through shells of Bergman (BG) type each two steps; here we call it BG icosahedral series;

2. the series of core-shell icosahedra with a Mackay (MC) core surrounded by a thick layer of shells of chiral-1 (Ch1) type, as are obtained in the growth of Ni@Ag and Cu@Ag core@shell nanoparticles (see Figure 5d-e in the main text).

### S1.2.1  The BG series

The pathway in the closed-packed plane corresponding to the BG icosahedral series is

$$(0,0) \to (0,1) \to (1,1) \to (1,2) \to (2,2) \to (2,3) \to ... \to (h,h) \to (h,h+1) \to ... \tag{S5}$$

The index $h$ is incremented by 1 every two steps, from 0 to $h_{max}$; the first and the second shell of index $h$ have $k = h$ and $k = h + 1$, respectively. We consider an icosahedron made of $i$ shells. If $i$ is even, the maximum value of $h$ is

$$h_{max} = \frac{i-2}{2}, \tag{S6}$$

and, in the last shell, $k = h_{max} + 1$. On the other hand, if $i$ is odd, we have

$$h_{max} = \frac{i-1}{2}, \tag{S7}$$

and, in the last shell, $k = h_{max}$. We calculate the total number of particles in the icosahedron, in both cases. For even $i$, we have

$$\begin{aligned} N_{EVEN} &= \sum_{h=0}^{h_{max}} \sum_{k=h}^{h+1} (10\, T^{(h,k)} + 2) - 1 = \\ &= \sum_{h=0}^{h_{max}} \sum_{k=h}^{h+1} \left[ 10\, (h^2 + k^2 + hk) + 2 \right] - 1. \end{aligned} \tag{S8}$$

Here we sum up the number of particles in each CK shell, up to $(h_{max}, h_{max} + 1)$. The sizes of the shells are given by Eq. S4, except for the first shell, namely $(0, 0)$; if we calculate the size of such shell by Eq. S4 we would obtain 2, which is wrong since the shell is made of only one particles. Therefore, if we use Eq. S4 for all shells, we have to subtract 1 particle from the final result. By writing explicitly the two terms $k = h$ and $k = h + 1$, we obtain

$$\begin{aligned} N_{EVEN} &= \sum_{h=0}^{h_{max}} (60\, h^2 + 30\, h + 14) - 1 = \\ &= 60 \sum_{h=0}^{h_{max}} h^2 + 30 \sum_{h=0}^{h_{max}} h + 14 \sum_{h=0}^{h_{max}} 1 - 1. \end{aligned} \tag{S9}$$

By using the equivalences

$$\begin{aligned} \sum_{h=0}^{h_{max}} h &= \frac{h_{max}(h_{max} + 1)}{2}, \\ \sum_{h=0}^{h_{max}} h^2 &= \frac{h_{max}(h_{max} + 1)(2h_{max} + 1)}{6}, \end{aligned} \tag{S10}$$

34

we calculate

$$N_{EVEN} = 20h_{max}^3 + 45h_{max}^2 + 39h_{max} + 13. \tag{S11}$$

Finally, we use the relation in Eq. S6 to write the total number of particles as function of the number of icosahedral shells

$$N_{EVEN}^{(i)} = \frac{5}{2}i^3 - \frac{15}{4}i^2 + \frac{9}{2}i - 1. \tag{S12}$$

If $i$ is odd, the CK indexes of the last shell are $(h_{max}, h_{max})$. The total number of particles in the icosahedron can be easily calculated, by subtracting from $N_{EVEN}$ the number of particles in the shell $(h_{max}, h_{max} + 1)$:

$$N_{ODD} = N_{EVEN} - N_{SHELL}^{(h_{max}, h_{max}+1)} =$$
$$= 20h_{max}^3 + 45h_{max}^2 + 39h_{max} + 13 - (30h_{max}^2 + 30h_{max} + 12) = \tag{S13}$$
$$= 20h_{max}^3 + 15h_{max}^2 + 9h_{max} + 1.$$

By using the relation in Eq. S7 we obtain

$$N_{ODD}^{(i)} = \frac{5}{2}i^3 - \frac{15}{4}i^2 + \frac{9}{2}i - \frac{9}{4}. \tag{8}$$

The number of particles in a icosahedron of the BG series made of $i$ shells is therefore

$$N_{BG}^{(i)} = \frac{1}{4}(10i^3 - 15i^2 + 18i - b), \tag{S14}$$

with

$$b = \begin{cases} 4 \text{ if } i \text{ is even} \\ 9 \text{ if } i \text{ is odd} \end{cases}. \tag{S15}$$

### S1.2.2 MC@Ch1 series

We consider an icosahedron made of $i$ MC shells, surrounded by $j$ Ch1 shells. The corresponding pathway in the closed-packed plane is

$$(0,0) \to (0,1) \to ... \to (0, i-1) \to (1, i-1) \to (1, i) \to ... \to (1, i+j-2) \tag{9}$$

35

The number of particles in the icosahedron is given by the sum of the number of particles in the MC core and the number of particles in the Ch1 shells. In the core, we have

$$N_{MC}^{(i)} = \sum_{k=0}^{i-1}(10\,T^{(0,k)} + 2) - 1 =$$
$$= \sum_{k=0}^{i-1}(10k^2 + 2) - 1 = \quad (S16)$$
$$= \frac{10}{3}i^3 - 5i^2 + \frac{11}{3}i - 1$$

particles, which is the well-known formula for MC icosahedral magic numbers. The total number of particles in the Ch1 shells is

$$N_{Ch1}^{(i,j)} = \sum_{k=i-1}^{i+j-2}(10\,T^{(1,k)} + 2) =$$
$$= \sum_{k=i-1}^{i+j-2}\left[10(k^2 + k + 1) + 2\right] = \quad (S17)$$
$$= \sum_{k=i-1}^{i+j-2}(10k^2 + 10k + 12).$$

In order to use the equivalences of Eq. S10, we introduce the index $l = k - i + 1$. We can write

$$N_{Ch1}^{(i,j)} = \sum_{l=0}^{j-1}\left[10(l + i - 1)^2 + 10(l + i - 1) + 12\right] =$$
$$= \sum_{l=0}^{j-1}\left[10l^2 + (20i - 10)l + 10i^2 - 10i + 12\right], \quad (S18)$$

and finally we calculate

$$N_{Ch1}^{(i,j)} = j\left(10i^2 + \frac{10}{3}j^2 + 10ij - 20i - 10j + \frac{56}{3}\right). \quad (S19)$$

The total number of particles in the MC@Ch1 icosahedron is therefore

$$N_{MC@Ch1}^{(i,j)} = N_{MC}^{(i)} + N_{Ch1}^{(i,j)}. \quad (S20)$$

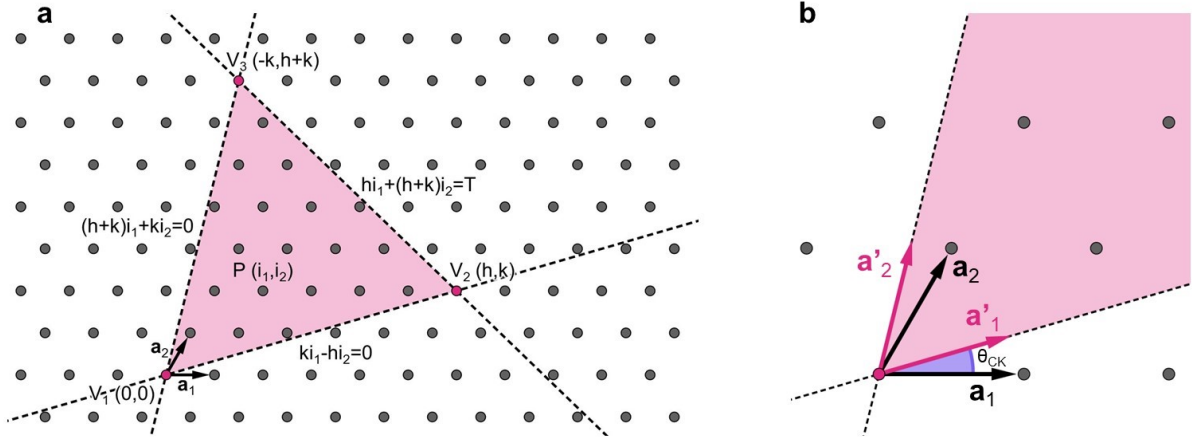

**Figure 8: Implementation of the CK construction.**
**a** Points belonging to the triangular face of the icosahedron are those enclosed by the three lines passing through the vertices, whose equations with respect to the basis primitive vectors $\mathbf{a}_1$ and $\mathbf{a}_2$ are reported in the figure. **b** Vectors $\mathbf{a}'_1$ and $\mathbf{a}'_2$, lying on two adjacent segment $V_1V_2$ and $V_1V_3$. $\mathbf{a}'_1$ and $\mathbf{a}'_2$ are obtained by rotating the primitive vectors $\mathbf{a}_1$ and $\mathbf{a}_2$ counterclockwise of an angle $\theta_{CK}$.

We can obtain different series of magic numbers by fixing the size of the MC core and adding Ch1 shells one after the other on top of it. Here we report the first magic numbers for some series of this kind.

- 2 MC shells: 1, 13, 45, 117, 249, 461, 773, 1205, ...

- 3 MC shells: 1, 13, 55, 127, 259, 471, 783, 1215, ...

- 4 MC shells: 1, 13, 55, 147, 279, 491, 803, 1235, ...

- 5 MC shells: 1, 13, 55, 147, 309, 521, 833, 1265, ...

- 6 MC shells: 1, 13, 55, 147, 309, 561, 873, 1305, ...

## S1.3   Implementation of the construction of a CK shell

Here we explain in details how the construction of CK shells is implemented in our C++ code. In the CK scheme, each icosahedral facet is an equilateral triangle whose vertices are lattice

points of a 2D hexagonal lattice. The coordinates of the vertices are given with respect to the the primitive vectors $\mathbf{a}_1$ and $\mathbf{a}_2$, forming an angle of $60°$ (see Supplementary Figure 8a): $V_1 = (0, 0)$, $V_2 = (h, k)$, $V_3 = (-k, h+k)$, where the positive integers $h$ and $k$ are the indexes of the CK construction, that unambiguously determine the icosahedral shell structure. All lattice points within the triangle belong to the CK shell, i.e. a lattice point $P = (i_1, i_2)$ belongs to the shell if $i_1$ and $i_2$ meet the following conditions:

$$\begin{cases} k\,i_1 - h\,i_2 \leq 0 \\ (h+k)\,i_1 + k\,i_2 \geq 0 \\ h\,i_1 + (h+k)\,i_2 \leq T \end{cases} \tag{S21}$$

where $T$ is the square distance between the vertices, i.e.

$$T = h^2 + k^2 + hk. \tag{S22}$$

It is useful to write the point $P$ with respect to the vectors $\mathbf{a}'_1$ and $\mathbf{a}'_2$, lying on two adjacent sides of the triangle. These are obtained by rotating the primitive vectors of the hexagonal lattice counterclockwise of an angle $\theta_{CK}$ (see Supplementary Figure 1b). The primitive vectors can be written as

$$\begin{cases} \mathbf{a}_1 = \left(\cos\theta_{CK} + \dfrac{\sin\theta_{CK}}{\sqrt{3}}\right) \mathbf{a}'_1 - \dfrac{2\sin\theta_{CK}}{\sqrt{3}} \mathbf{a}'_2 \\ \mathbf{a}_2 = \dfrac{2\sin\theta_{CK}}{\sqrt{3}} \mathbf{a}'_1 + \left(\cos\theta_{CK} - \dfrac{\sin\theta_{CK}}{\sqrt{3}}\right) \mathbf{a}'_2 \end{cases} \tag{S23}$$

The cosine and sine of $\theta_{CK}$ are calculated from the CK indexes, as

$$\cos\theta_{CK} = \frac{2h+k}{2\sqrt{T}}, \tag{S24}$$

$$\sin\theta_{CK} = \frac{\sqrt{3}\,k}{2\sqrt{T}}. \tag{S25}$$

The coordinates of the lattice point $P$ can therefore be expressed as

$$P - V_1 = \left\{ i_1 \cos\theta_{CK} + (i_1 + 2i_2)\frac{\sin\theta_{CK}}{\sqrt{3}} \right\} \mathbf{a}'_1 + \left\{ i_2 \cos\theta_{CK} - (2i_1 + i_2)\frac{\sin\theta_{CK}}{\sqrt{3}} \right\} \mathbf{a_2}'. \tag{S26}$$

For building a CK icosahedral shell by our C++ code, the CK indexes $h$ and $k$ and the inter-particle distance $d_{CK}$ have to be set by user. The quantities $T$, $\cos\theta_{CK}$ and $\sin\theta_{CK}$ are calculated, then the CK is built in according to the following scheme:

1. The first 12 particles are placed in the vertices of the icosahedron of edge length $d_{CK}\sqrt{T}$.

2. For each triplet of vertices $V_1$, $V_2$, $V_3$ forming an icosahedral facet, the vectors $\mathbf{a}'_1$ and $\mathbf{a_2}'$ are calculated as

$$\mathbf{a}'_1 = \frac{V_2 - V_1}{\sqrt{T}}, \tag{S27}$$

$$\mathbf{a}'_2 = \frac{V_3 - V_1}{\sqrt{T}}. \tag{S28}$$

3. Particles within the icosahedral facet are placed on points with coordinates given by Eq. S26, for all couples of indexes $i_1$, $i_2$ meeting the conditions of Eq. S21.

## S1.4 Evaluation of the optimal size mismatch in CK icosahedra

Here we discuss the evaluation of the optimal size mismatch between particles in different shells of a multi-shell CK icosahedron. We start by evaluating the optimal size mismatch in core-shell icosahedra made of a MC core surrounded by one shell of either MC or Ch1 type (namely MC@MC and MC@Ch1 icosahedra); then, we extend our results to structures made of CK shells belonging to any chirality class.

### S1.4.1 Optimal size mismatch in the MC@MC icosahedron

In MC shells, one of the CK indexes is equal to zero. Here we consider icosahedral shells with $k \geq h$; therefore MC shells have $h = 0$, whereas $k$ can take any integer value. The triangulation number is

$$T_{MC}^{(k)} = k^2, \tag{10}$$

therefore the radius of the shell is

$$r_{MC}^{(k)} = d_{CK} \sin\left(\frac{2\pi}{5}\right) k, \tag{S29}$$

$d_{CK}$ being the inter-particle distance in the 2D hexagonal plane. The MC icosahedron is made of concentric MC shells. Let us now calculate *inter-shell* distances, i.e. distances between icosahedral vertices belonging to nearby shells. This corresponds to the difference between the radii

$$\begin{aligned}\Delta r = r_{MC}^{(k+1)} - r_{MC}^{(k)} = \\ = (k+1)\, d_{CK} \sin\left(\frac{2\pi}{5}\right) - k\, d_{CK} \sin\left(\frac{2\pi}{5}\right) = \\ = d_{CK} \sin\left(\frac{2\pi}{5}\right) \approx 0.9511\, d_{CK}\end{aligned} \tag{S30}$$

The inter-shell distance is the same for all pairs of nearby shells, and it is smaller than the intra-shell distance by about 5%. It is possible to verify that all nearest-neighbours distances between particles in the same shell are equal to $d_{CK}$, and those between particles in nearby shells are equal to $\Delta r$; therefore, in the MC icosahedron, the inter-particle distance spectrum has only two peaks, in $\Delta r$ and $d_{CK}$.

Let us now consider a MC icosahedron made of identical spherical particles. We denote by $d$ the ideal pair distance of the system. In the case of metal atoms, $d$ corresponds to the ideal interatomic distance in the bulk crystal. Since intra-shell and inter-shell distances within the MC icosahedron are different, it is not possible to build a structure of this kind in which all pair distances are equal to $d$. However, we expect the icosahedron to be stable if pair distances are close enough to the ideal value. We want to quantify this concept, i.e. to determine what is the optimal MC icosahedral arrangement when packing particles of the same type. Such optimal arrangement depends on the ability of the particles (spheres, atoms, molecules, ...) to adapt pair distances, by expanding or compressing them from the ideal value. Here we consider a system in which pair distances can contract and expand symmetrically around the ideal distance. This

model should describe reasonably well the behaviour of atomic systems. In such a system, $\Delta r$ and $d_{CK}$ are expected to be displaced from $d$ by the same amount, i.e.

$$d_{CK} = (1 + \xi) d, \tag{S31}$$

$$\Delta r = (1 - \xi) d. \tag{S32}$$

By using the relation between inter- and intra-shell distances in Eq. S30, we find the optimal expansion/contraction coefficient $\xi_{MC}$ of pair distances for a MC icosahedron made of spherical particles of a single type

$$\xi_{MC} = \frac{1 - \sin\left(\frac{2\pi}{5}\right)}{1 + \sin\left(\frac{2\pi}{5}\right)} \approx 0.0251. \tag{S33}$$

Now we consider bi-elemental icosahedra of core-shell type. The icosahedron is made of two parts: an internal core, which is an icosahedron made of multiple concentric MC shells, and a single external MC shell, of a different element. Ideal pair distances in the core and in the shell are $d_{CORE}$ and $d_{SHELL}$, respectively. We denote by $k$ the index of the most external shell of the core; the index of the shell is therefore $k + 1$. For the core, we consider the optimal MC icosahedral arrangement, in which intra-shell distances are expanded by $\xi_{MC}$ compared to the ideal distance $d_{CORE}$. The radius of the core is given by Eq. S29, and it is

$$r_{CORE} = k \sin\left(\frac{2\pi}{5}\right)(1 + \xi_{MC}) d_{CORE}. \tag{S34}$$

Since the shell is made of one single MC layer, all pair distances within the shell are equal; in the optimal shell, such distances are equal to the ideal value $d_{SHELL}$. The radius of the shell is therefore

$$r_{SHELL} = (k + 1) \sin\left(\frac{2\pi}{5}\right) d_{SHELL}. \tag{S35}$$

The core-shell distance is given by

$$\Delta r_{CORE-SHELL} = r_{SHELL} - r_{CORE} =$$
$$= (k + 1) \sin\left(\frac{2\pi}{5}\right) d_{SHELL} - k \sin\left(\frac{2\pi}{5}\right)(1 + \xi_{MC}) d_{CORE}, \tag{S36}$$

In the optimal core-shell icosahedron, the core-shell distance exactly corresponds to the ideal value

$$d_{CORE-SHELL} = \frac{d_{CORE} + d_{SHELL}}{2}. \tag{S37}$$

In this way, the pair distance distribution within the whole structure is as close as possible to the ideal interatomic distances $d_{CORE}$, $d_{SHELL}$ and $d_{CORE-SHELL}$. We note that $\Delta r$ corresponds to the distance between particles in nearby vertex of adjacent shells. If such shells are made of the same atom type, all inter-shell nearest-neighbour distances are equal to $\Delta r$, but this is not true for shells made of particles of different sizes, in which a more complex nearest-neighbour distance spectrum arises. The condition $\Delta r = d_{CORE-SHELL}$ is therefore to be taken as an approximate condition, which, as we will see in the following, allows to establish an approximate relationship between $d_{CORE}$ and $d_{SHELL}$.

The condition on $\Delta r$ translates into the following relation between $d_{CORE}$ and $d_{SHELL}$:

$$(k+1)\sin\left(\frac{2\pi}{5}\right) d_{SHELL} - k \sin\left(\frac{2\pi}{5}\right)(1+\xi_{MC}) d_{CORE} = \frac{d_{CORE} + d_{SHELL}}{2}. \tag{S38}$$

We define the ratio

$$\alpha = \frac{d_{SHELL}}{d_{CORE}} \tag{S39}$$

between ideal interatomic distances in the core and in the shell. Eq. S38 can be written in terms of $\alpha$, as

$$(k+1)\sin\left(\frac{2\pi}{5}\right)\alpha - k\sin\left(\frac{2\pi}{5}\right)(1+\xi_{MC}) = \frac{1+\alpha}{2}. \tag{S40}$$

We can therefore find the ideal value $\alpha_{MC-MC}$ for which the optimal core-shell MC@MC icosahedron can be built, as a function of the index $k$ of the most external MC shell of the core:

$$\alpha_{MC-MC}^{(k)} = \frac{2(1+\xi_{MC})\sin\left(\frac{2\pi}{5}\right)k + 1}{2\sin\left(\frac{2\pi}{5}\right)(k+1) - 1}. \tag{S41}$$

We define the size mismatch between the elements in the core and in the shell

$$sm = \frac{d_{SHELL} - d_{CORE}}{d_{CORE}} = \alpha - 1. \tag{S42}$$

42

The ideal size mismatch is

$$sm_{MC-MC}^{(k)} = \alpha_{MC-MC}^{(k)} - 1. \tag{S43}$$

We note that, if we start from a MC icosahedral core, and we add a further layer made of particles of the same type, we have to adapt inter-particle distances of the newly added layer according to the optimal distance of Eq. S31, which is larger compared to the ideal one. The inter-shell distance between the core and the new layer does not correspond to the ideal distance, as well. If we use particles of a different size for building the shell, it is possible to have a perfect match, in which both intra-shell distances within the shell and inter-shell distances between the core and the shell exactly correspond to the ideal ones. Of course we have to select particles of the right size, according to Eq. S41. In this way, we expect the stability of the icosahedron to be improved.

### S1.4.2 Optimal size mismatch in the MC@Ch1 icosahedron

Here we consider core-shell icosahedra of a different type. The internal core is a complete MC icosahedron, whereas the external shell is of Ch1 type. Again, we consider bi-elemental structures, in which the core and the shell are made of different particles, with ideal pair distances $d_{CORE}$ and $d_{SHELL}$. The CK indexes of the most external shell of the MC core are $(0, k)$, whereas the CK indexes of the shell are $(1, k)$: passing from the core to the shell we increment the index $h$ from 0 to 1, while the index $k$ is unchanged (here we follow the rules for building CK icosahedra that we have discussed in the main text). The radius of the core is

$$r_{CORE} = k \sin\left(\frac{2\pi}{5}\right)(1 + \xi_{MC}) d_{CORE}, \tag{S44}$$

where $\xi_{MC}$ is given in Eq. S33. The triangulation number of the $(1, k)$ Ch1 shells is

$$T_{Ch1}^{(k)} = k^2 + k + 1, \tag{S45}$$

and therefore the shell radius is

$$r^*_{SHELL} = d_{SHELL} \sin\left(\frac{2\pi}{5}\right)\sqrt{k^2 + k + 1}. \tag{S46}$$

We note that it is not possible to keep all inter-particle distances equal to $d_{SHELL}$ in the Ch1 shell. Intra-facet distances are equal to $d_{SHELL}$, but inter-facet distances are smaller (here, a facet is a triangle formed by three nearby icosahedral vertices). Indeed, this occurs in all CK shells, except for the MC ones. Even though not all nearest-neighbour distances within the Ch1 shell are equal to the ideal one, the great majority of the pair distances are of intra-facet kind, and are therefore equal to $d_{SHELL}$. Therefore, the arrangement that we are considering is a reasonably good estimate of the optimal shell arrangement; this is true expecially for large shells, in which the ratio between the number of intra-facet and inter-facet distances is large.

We calculate the core-shell distance $\Delta r$, and we impose

$$\Delta r = \frac{d_{CORE} + d_{SHELL}}{2}. \tag{S47}$$

In this way, we find the condition for building the optimal core-shell MC@Ch1 icosahedron

$$\alpha^{(k)}_{MC-Ch1} = \frac{2(1 + \xi_{MC})\sin\left(\frac{2\pi}{5}\right)k + 1}{2\sin\left(\frac{2\pi}{5}\right)\sqrt{k^2 + k + 1} - 1}. \tag{S48}$$

where $\alpha$ is the ratio between $d_{SHELL}$ and $d_{CORE}$. The optimal size-mismatch between the MC core and the Ch1 shell can be calculated as

$$sm^{(k)}_{MC-Ch1} = \alpha^{(k)}_{MC-Ch1} - 1. \tag{S49}$$

### S1.4.3 General formula for the evaluation of the optimal size mismatch

The formula in Eq. S41 and S48 can be generalized for estimating the optimal size mismatch at any interface between icosahedral shells. We consider a core made of concentric icosahedral shells. Unlike the previous case of a mono-elemental MC core, here the shells building up the

44

core the can be different, i.e. they can belong to different chirality classes, and can be made of particles of different sizes. We only assume that all particles within the same shell are equal. We denote by $d_{CORE}$ the ideal distance of the element in the most external shell of the core. On top of the core, we put a further shell, made of particles with ideal pair distance $d_{SHELL}$. The radius of the core is

$$r_{CORE} = \sin\left(\frac{2\pi}{5}\right)(1+\xi)\sqrt{T_{CORE}}\, d_{CORE}, \tag{S50}$$

where $T_{CORE}$ is the triangulation number of the most external shell of the core. As for the mono-elemental MC core, we put the expansion coefficient $\xi$ of the intra-facet distances of the icosahedral layer. The evaluation of $\xi$ will be discussed in the following.

The radius of the shell is

$$r_{SHELL} = d_{SHELL} \sin\left(\frac{2\pi}{5}\right)\sqrt{T_{SHELL}}, \tag{S51}$$

where $T_{SHELL}$ is the triangulation number of the shell. For the shell, we consider $d_{CK} = d_{SHELL}$. In the case of MC shells, all inter-shell distances are equal to $d_{SHELL}$; in all other cases, only intra-facet distances are equal to $d_{SHELL}$, but, since they are the majority, our choice is expected to be close to the optimal shell arrangement.

We calculate the core-shell distance $\Delta r$, and we impose

$$\Delta r = \frac{d_{CORE} + d_{SHELL}}{2}. \tag{S52}$$

In this way, we find the condition for the optimal core-shell interface

$$\alpha = \frac{2(1+\xi)\sin\left(\frac{2\pi}{5}\right)\sqrt{T_{CORE}}+1}{2\sin\left(\frac{2\pi}{5}\right)\sqrt{T_{SHELL}}-1}. \tag{S53}$$

The optimal size mismatch can be calculated as

$$sm = \alpha - 1. \tag{S54}$$

Let us now discuss the evaluation of the expansion coefficient $\xi$ of intra-shell distances in the most external shell of the core. The expansion arises when multiple icosahedral shells of the same element are present in the structure. Keeping the same element when passing from one shell to the outer one is never the optimal choice, since the optimal size mismatch is always larger than zero (even in the MC case, which is the one with the smallest size mismatch between nearby shells). As a consequence, if in the icosahedron there is a thick layer made of multiple shells of the same element, such layer will undergo some expansion of inter-shell distances in order to bring the the nearest-neighbour distance spectrum closer to the ideal value. We can distinguish two cases:

1. the most external shell of the core is isolated, i.e. it is the only one made of that specific particle type; the inner shells are made of different elements;

2. the most external shell of the core is the most external shell of a thick layer made of multiple shells of the same element (at least two shells).

In the first case, $\xi$ is equal to zero. Since we are building optimal icosahedral structures, atoms in the most external layer of the core are those with the optimal size-mismatch with the underlying shell. The inter-shell distance is the optimal one, and no expansion is therefore necessary.

Let us now analyse the second case. We consider a thick layer made of $n$ shells of the same element, with ideal pair distance $d_{CORE}$. The first shell is on top of an icosahedral shell of a different element, and the distance $d_{CORE}$ is chosen accordingly to the size-mismatch at such inner interface; therefore, inter-shell distances in the first shell exactly correspond to $d_{CORE}$ and the radius is

$$r_1 = d_{CORE} \sin\left(\frac{2\pi}{5}\right) \sqrt{T_1}, \tag{S55}$$

where $T_1$ is the triangulation number of the first shell of the layer. The radius of the second shell

is

$$r_2 = d_2 \sin\left(\frac{2\pi}{5}\right) \sqrt{T_2}, \tag{S56}$$

where $T_2$ is the triangulation number of the shell, and $d_2$ is the inter-shell distance within the shell. Here we want to determine the optimal value of $d_2$. We calculate the inter-shell distance

$$\Delta r_{1,2} = r_2 - r_1 = d_2 \sin\left(\frac{2\pi}{5}\right) \sqrt{T_2} - d_{CORE} \sin\left(\frac{2\pi}{5}\right) \sqrt{T_1}. \tag{S57}$$

Since we are considering systems with symmetric expansion/contraction from the ideal distance, we assume that the optimal inter- and intra-shell distances are those satisfying the relation

$$d_{CORE} - \Delta r_{1,2} = d_2 - d_{CORE}. \tag{S58}$$

We use the relation between $\Delta r_{12}$ and $d_2$ in Eq. S57, and we obtain

$$d_2 = \frac{2 + \sin\left(\frac{2\pi}{5}\right)\sqrt{T_1}}{1 + \sin\left(\frac{2\pi}{5}\right)\sqrt{T_2}} d_{CORE}. \tag{S59}$$

We define the expansion coefficient $\xi_2$ in the second shell as

$$d_2 = (1 + \xi_2) d_{CORE}. \tag{S60}$$

We therefore calculate

$$\xi_2 = \frac{1 - \sin\left(\frac{2\pi}{5}\right)(\sqrt{T_2} - \sqrt{T_1})}{1 + \sin\left(\frac{2\pi}{5}\right)\sqrt{T_2}}. \tag{S61}$$

We can follow the same procedure for evaluating the expansion coefficient of the outer shells, up to the most external one. We consider the interface between the $i$-th shell and the $(i+1)$-th shell, $1 \leq i \leq n-1$. The corresponding radii are

$$\begin{aligned} r_i &= d_i \sin\left(\frac{2\pi}{5}\right) \sqrt{T_i} \\ r_{i+1} &= d_{i+1} \sin\left(\frac{2\pi}{5}\right) \sqrt{T_{i+1}}, \end{aligned} \tag{S62}$$

from which we calculate the inter-shell distance

$$\Delta r_{i,i+1} = r_{i+1} - r_i = d_{i+1} \sin\left(\frac{2\pi}{5}\right) \sqrt{T_{i+1}} - d_i \sin\left(\frac{2\pi}{5}\right) \sqrt{T_i}. \tag{S63}$$

Here we assume that the inter-shell distance in the $i$-th shell is known, as it has been determined in the previous step of this recursive procedure. Specifically, we have

$$d_i = (1 + \xi_i) \, d_{CORE} \tag{S64}$$

We impose the relation

$$d_{CORE} - \Delta r_{i,i+1} = d_{i+1} - d_{CORE}, \tag{S65}$$

from which we calculate

$$d_{i+1} = \frac{2 + (1 + \xi_i) \sin\left(\frac{2\pi}{5}\right) \sqrt{T_i}}{1 + \sin\left(\frac{2\pi}{5}\right) \sqrt{T_{i+1}}}, \tag{S66}$$

and finally

$$\xi_{i+1} = \frac{1 - \sin\left(\frac{2\pi}{5}\right) \left[\sqrt{T_{i+1}} - (1 + \xi_i)\sqrt{T_i}\right]}{1 + \sin\left(\frac{2\pi}{5}\right) \sqrt{T_{i+1}}}. \tag{S67}$$

We start from $i = 1$, for which we always have $\xi_1 = 0$, as the inter-shell distance in the first shell exactly corresponds to the ideal distance $d_{CORE}$; we calculate $\xi_2$, then $\xi_3$ and so on, up to $\xi_n$, i.e. the expansion coefficient in the most external shell of the icosahedral core, which has to be put in Eq. S53.

If we consider the case of a single-element MC core and we use Eq. S67 for calculating the expansion coefficients starting from the central atom, we find that the expansion coefficient is the same for each MC shell, and it is equal to $\xi_{MC}$, as expected.

By combining the formula in Eq. S53 and S67, we can estimate the optimal size mismatch at any interface between concentric CK shells. Some meaningful examples are reported in the main text.

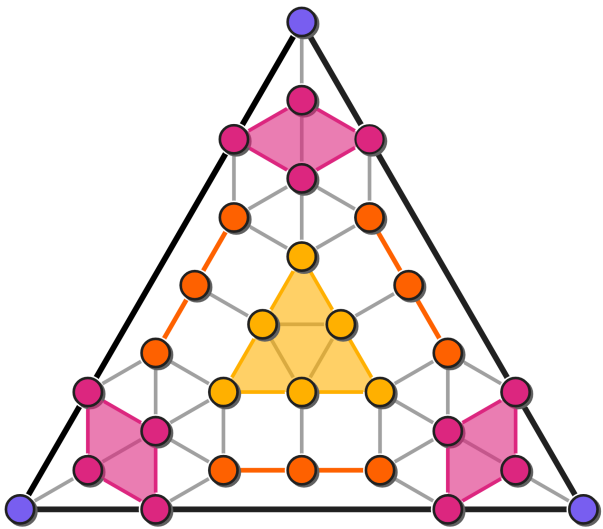

**Figure 9: Facet of an anti-Mackay shell.**
Particles are marked in different colours to help in counting the number of particles in the facet.

## S1.5 Anti-Mackay shells

Generalised anti-Mackay (AM) shells are non-closed-packed, achiral icosahedral shells. Each AM shell is identified univocally by the two AM indexes $p$ and $q$: $p$ is the number of particles on the side of the closed-packed equilateral triangle concentric to the icosahedral facet; $q$ is the number of particles on each median of the icosahedral facet, between the vertex of the icosahedral facet and the vertex of the inner triangle (see Figure 6a in the main text). As for the CK shells, simple geometric considerations allow us to identify the rules for packing AM shells into icosahedra, and to associate the optimal size mismatch to each possible interface between nearby shells of different kind.

Here we start by calculating the number of particle in a AM shell, as a function of the two AM indexes $p$ and $q$. It is useful to divide the particles of the icosahedral facet into different groups, as shown in Supplementary Figure 9:

1. Particles in the closed-packed inner triangle are marked in yellow in the figure. There are

$p$ particles on the side of the triangle, and therefore there are

$$N_T = \frac{p^2 + p}{2} \tag{S68}$$

particles in the triangle. In this case, $p = 3$ and therefore $N_T = 6$.

2. Particles forming square sites (not belonging to the inner triangle) are marked in orange in the figure. Three groups of this kind are present in the facet. Each group is a rectangle of sides $p$ and $q/2$, and therefore it has

$$N_S = \frac{pq}{2} \tag{S69}$$

particles. In this case, $p = 3$ and $q = 2$, therefore there are 3 particles in each group. We note that, if $p$ and $q$ are both odd, we obtain a semi-integer number. This occurs when a row of particles is on the edge of the icosahedral facet; such particles are shared between two facets, and therefore the contribution of each particle is 1/2.

3. The remaining particles are marked in pink in the figure. There are three groups, one for each vertex. Each group includes the $q$ particles on the median of the triangle, and other particles placed according to the hexagonal lattice pattern. The shape of each group is a diamond of edge $q$, truncated by the edges of the icosahedral facet. The sum of the two truncations is equivalent to an equilateral triangle of side $q - 1$, so that the number of particles in each diamond group is

$$N_D = q^2 - \frac{(q-1)^2 + q - 1}{2} = \frac{q^2 + q}{2}. \tag{S70}$$

In this case, there are 3 particles in each group; two of the pink particles in each group are on the edges of the icosahedral facet, therefore they are counted has half particles.

The total number of particles in the AM facet is therefore

$$N_{AM}^{facet} = N_T + 3N_S + 3N_D = \frac{p^2 + p + 3pq + 3(q^2 + q)}{2}. \tag{S71}$$

To obtain the total number of particles in the AM shell, we simply have to multiply $N_{AM}^{facet}$ by 20, and add the 12 particles in the vertices of the icosahedron:

$$N_{AM}^{(p,q)} = 10\left(p^2 + 3q^2 + 3pq + p + 3q\right) + 12, \tag{S72}$$

which can be written as

$$N_{AM}^{(p,q)} = 10\left(p^2 + 3q^2 + 3pq + p + 3q + 1\right) + 2. \tag{S73}$$

This expression looks similar to the formula for the number number of particles in a CK shell, i.e.

$$N_{CK}^{(h,k)} = 10T^{(h,k)} + 2 = 10(h^2 + k^2 + hk) + 2, \tag{S74}$$

where $h$ and $k$ are the indexes of the CK construction. Indeed, if we consider the CK shell of indexes

$$\begin{aligned} h &= q + 1, \\ k &= p + q, \end{aligned} \tag{S75}$$

we calculate

$$\begin{aligned} N_{CK}^{(q+1,p+q)} &= 10\left[(q+1)^2 + (p+q)^2 + (q+1)(p+q)\right] + 2 = \\ &= 10\left(p^2 + 3q^2 + 3pq + p + 3q + 1\right) + 2 = \\ &= N_{AM}^{(p,q)}, \end{aligned} \tag{S76}$$

so that each AM shell can be put in relation with a CK shell of the same size. More precisely, the AM shell can transform into the corresponding CK shell by some concerted rotations of the particles, with different rotation axes (perpendicular to the planes of the icosahedral facets) and different rotation angles and, viceversa, CK shells can transform into AM ones. In Figure 6b-c of the main text we have shown two examples:

- Ch1→AM1. In Ch1 shells, we have $h = 1$ (we consider shells with $k > h$) and therefore the corresponding AM shell has $q = 0$, i.e. no particles on the medians of the triangular

facet. The Ch1 facet transforms into the corresponding AM1 facet by a clockwise rotation of all particles (excluding the vertices) around the center of the facet. The rotation angle is $60° − \theta_{CK}$, where $\theta_{CK}$ is the angle of the CK construction, given by Eq. S24, S25. Ch1 shells with $k = 1$, $k < h$ transform into the same AM1 shells, by a counterclockwise rotation of $\theta_{CK}$.

- Ch2→AM2. In Ch2 shells, we have $h = 2$ (we consider shells with $k > h$) and therefore the corresponding AM shell has $q = 1$, i.e. one particle on each median of the triangular facet. The Ch2→AM2 transformation is more complicated, with different rotation angles and axes for different groups of particles in the icosahedral facet:

  1. particles in the inner triangle (marked in green in Figure 6c of the main text) rotate clockwise around the center of the facet of an angle $60° − \theta_{CK}$, as in the Ch1→AM1 transformation;

  2. particles close to the vertices (marked in blue in the figure) rotate around the closest vertex counterclockwise of an angle $\theta_{CK} − 30°$; in this way, they reach the medians of the icosahedral facet;

  3. the remaining particles (marked in yellow in the figure) rotate clockwise around the center of the closest facet edge, of an angle $60° − \theta_{CK}$; in this way, they reach the edge of the facet.

  Ch2 shells with $k = 2$, $k < h$ transform into the same AM2 shells; the transformation is of the same type, with opposite rotation directions, and rotation angles of $\theta_{CK}$ instead of $60° − \theta_{CK}$, and $30° − \theta_{CK}$ instead of $\theta_{CK} − 30°$.

We note that such transformations do not allow to obtain the optimal AM arrangement, since neighbouring particles rotating around different axes end up with inter-particle distances that

are shorter than the ideal one. This occurs because the radii of the perfect Ch and AM shells with the same number of particles are different, as we show in the following.

The edge of the AM shell with unit inter-particle distance and AM indexes $p$, $q$ can be easily calculated by looking at Supplementary Figure 9. It corresponds to the side of the inner triangle, which is $p-1$, plus twice the projection of the segment between the vertices of the facet and of the inner triangle (which is $q+1$ long) on the facet edge:

$$l_{AM}^{(p,q)} = p - 1 + 2(q+1)\frac{\sqrt{3}}{2} = p + \sqrt{3}q + \sqrt{3} - 1. \tag{S77}$$

We write the edge length as a function of the CK indexes of the corresponding CK shell, by substituting $p = k - h + 1$ and $q = h - 1$:

$$l_{AM}^{(h,k)} = (\sqrt{3} - 1)h + k, \tag{S78}$$

We recall that the edge length of the CK shell is

$$l_{CK}^{(h,k)} = \sqrt{T^{(h,k)}} = \sqrt{h^2 + k^2 + hk}. \tag{S79}$$

It is easy to verify that, for each couple of indexes $(h, k)$, $l_{AM}^{(h,k)} > l_{CK}^{(h,k)}$. The same holds for the radii, which are calculated by multiplying the edge length by $\sin(2\pi/5)$. CK shells are more compact than the corresponding AM ones, as particles are more densely packed: in CK shells, all particles are closed-packed (we recall that CK shells are obtained by folding an hexagonal closed-packed lattice), whereas in AM shells only a portion of particles is in closed-packed arrangement, the other ones forming only 5 or 4 nearest-neighbours bonds.

We can estimate the optimal size mismatch for icosahedra with AM shells by using the same geometric consideration made for CK icosahedra. Indeed, we can use the formula in Eq. S53 and S67, in which we shall replace $\sqrt{T}$ by the length of the AM edge in Eq. S78. As an example, we calculate the optimal size mismatch for adding a AM1 shell on a MC core, i.e.

between the $(0, k)$ and $(1, k)^*$ icosahedral shells, with $k \geq 1$. The length of the edge of the $(1, k)^*$ shell is

$$l_{AM}^{(1,k)} = k + \sqrt{3} - 1, \tag{S80}$$

and therefore the optimal size mismatch is given by

$$sm_{MC-AM1}^{(k)} = \frac{2(1 + \varepsilon_{MC})\sin\left(\frac{2\pi}{5}\right)k + 1}{2\sin\left(\frac{2\pi}{5}\right)(k + \sqrt{3} - 1) - 1} - 1. \tag{S81}$$

Values are reported in Figure 7a of the main text.

## S1.6 Comparison of the stability of AM1 and Ch1 shells

The data on the optimal mismatch in Lennard-Jones and Morse clusters have been obtained by calculating the energy, after local minimization, as a function of the mismatch for the different types of structures, and looking for the mismatch at which the energy is minimum. Two examples are reported in Supplementary Figure 10, where we consider MC, AM1 and Ch1 shells on Mackay cores made of 5 and 8 MC shells. The MC structures are better stabilised in a low-mismatch range. The AM1 shells find their optimal mismatch in an intermediate range and after a threshold mismatch they become unstable, i.e. they are not even local minima. On the contrary, the Ch1 structures are stable in a high-mismatch range and become unstable below a threshold mismatch. In the cases reported in Supplementary Figure 10 the range of mismatch values in which both AM1 and Ch1 structures are local minima is very narrow. In Supplementary Figure 10a we have reported the same data of Figure 3a of the main text for the addition of either a MC or a Ch1 shell on a five-shell Mackay core; data for the addition of a AM1 shell have been included, in order to compare the three possible arrangements of the outer shell.

## S1.7 Packing of AM shells

In the main text we have evaluated the optimal mismatch for adding a $(1, k)^*$ AM1 shell on a MC core terminated by a $(0, k)$ shell. Here we consider the possibility of adding a further shell

54

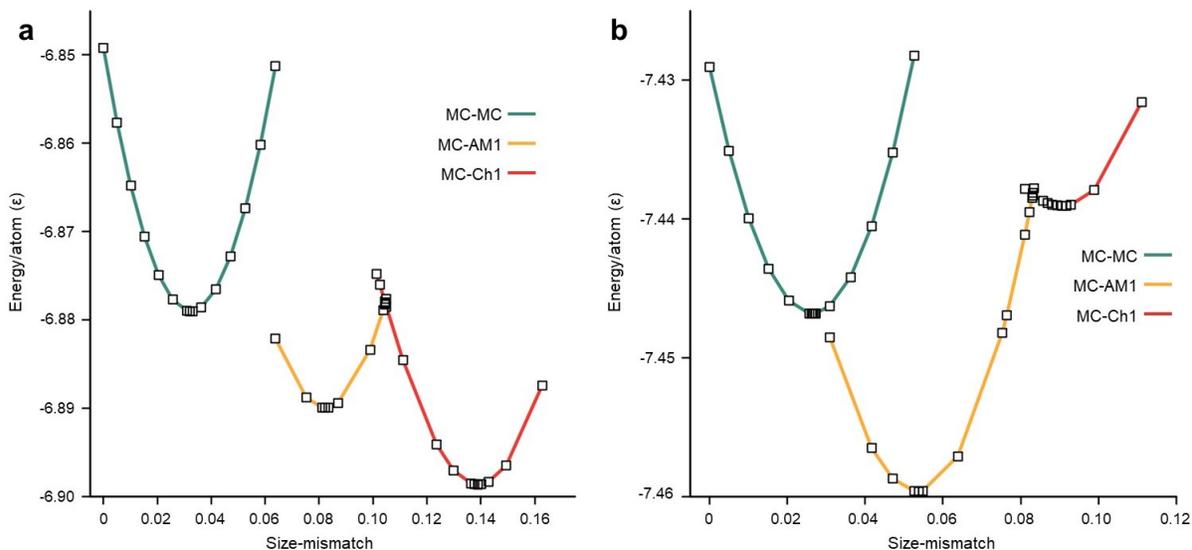

**Figure 10: Stability of MC, AM1 and Ch1 shells in Lennard-Jones clusters.**
Binding energy per particle (in $\varepsilon$ units) as a function of the size mismatch of **a** an icosahedron of $i = 5$ MC shells covered by single MC, AM1 and Ch1 shells and **b** an icosahedron of $i = 8$ MC shells covered by single MC, AM1 and Ch1 shells.

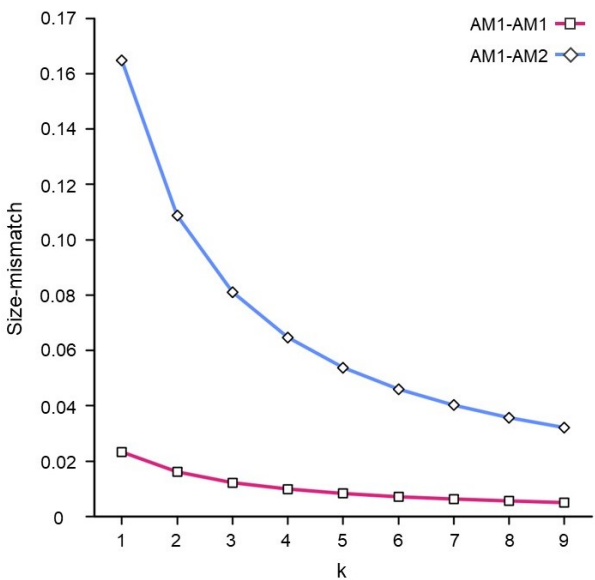

**Figure 11: Optimal mismatch for two AM shells on a MC core.**
A core terminated by a $(1, k)^*$ AM1 shell is covered by either a $(1, k+1)^*$ or a $(2, k)^*$ shell. The optimal mismatch is evaluated according to Eq. S82 and Eq. S83.

55

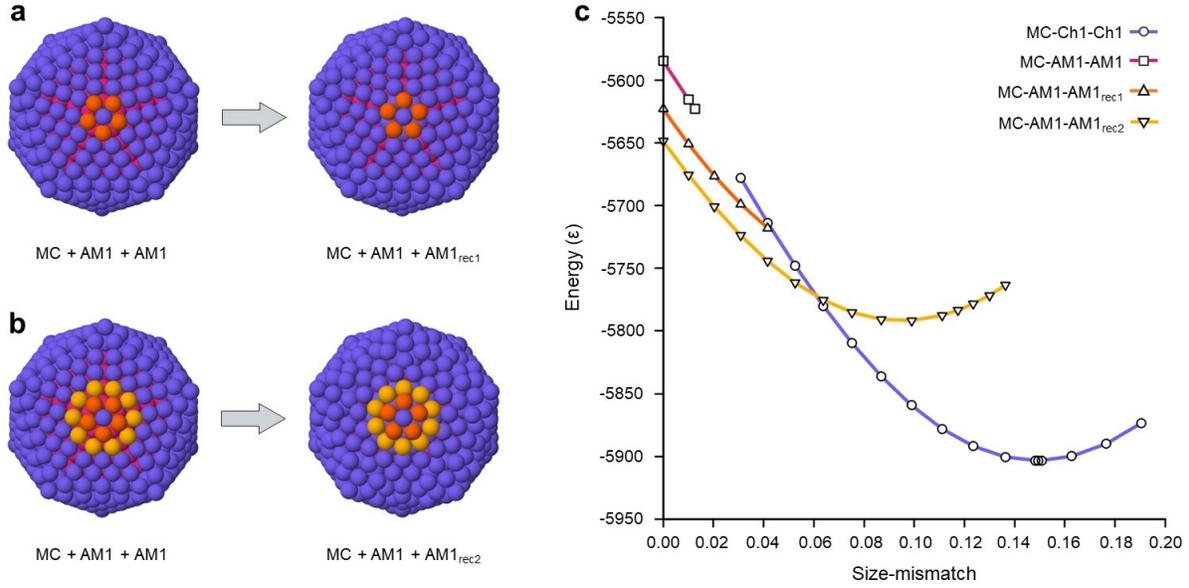

**Figure 12: Stability of two AM shells on a MC core.**
**a-b** The two types of reconstruction undergone by the outer AM1 shell spontaneously upon local minimization, involving rotations of particle rings around all vertices. **c** Energy (in $\varepsilon$ units) as a function of the mismatch for Lennard-Jones clusters made of 5 Mackay shells upon which two shells are added: two Ch1 shells; two AM1 shells; an AM1 shell plus a shell reconstructed as in **a**; an AM1 shell plus a shell reconstructed as in **b**.

of AM type. According to the path rules, the shell can be either of $(1, k+1)^*$ or $(2, k)^*$ type, i.e. AM1 or AM2. The optimal mismatch can be analytically evaluated as explained in the previous Sections, obtaining the formula below:

$$sm_{AM1-AM1}^{(k)} = \frac{2\sin\left(\frac{2\pi}{5}\right)(k+\sqrt{3}-1)+1}{2\sin\left(\frac{2\pi}{5}\right)(k+\sqrt{3})-1} - 1, \tag{S82}$$

$$sm_{AM1-AM2}^{(k)} = \frac{2\sin\left(\frac{2\pi}{5}\right)(k+\sqrt{3}-1)+1}{2\sin\left(\frac{2\pi}{5}\right)(k+2\sqrt{3})-2} - 1. \tag{S83}$$

Values of the optimal mismatch are reported in Supplementary Figure 11. As in the case of chiral shells (see Figure 3d in the main text), the optimal mismatch is larger for the class-changing step $(1, k)^* \to (2, k)^*$. This may lead to the conclusion that the outer shell should be preferentially of AM1 type (i.e. $(1, k+1)^*$) if the mismatch is zero or low. But this is not the case, as we show in Supplementary Figure 12. In fact the mismatch range in which the configuration with two AM1 shells is a local minimum is very narrow, and even when locally

56

stable, its energy is quite high. Outside that range, the outer shell spontaneously reconstructs upon local minimization, as shown in Supplementary Figure 12a-b. This is due to the very unfavourable coordination of particles on the facet edges of the outer shell, which sit on only two nearest neighbours of the shell below (this point will be further discussed in the last Section). Both reconstructions improve that coordination by displacing atoms on fourfold sites of the substrate.

In Supplementary Figure 12c we calculate the energies of unreconstructed and reconstructed configurations of Lennard-Jones clusters. The reconstruction of Supplementary Figure 12b is more favourable, as it displaces a larger number of particles on sites of higher coordination with the substrate. Due to the instability of the configuration with two AM1 shells, the natural growth sequences lead to the formation of an outer AM2 shell, as shown in Figure 7b-c of the main text and discussed in the Supplementary Note 3.

Finally, we note that the configuration with two Ch1 shells, which has the same number of particles, is much more favourable in a wide range of mismatch, as shown in Supplementary Figure 12c. In fact, from the results in Supplementary Figure 12c it turns out also that the stability range of two Ch1 shells is enlarged with respect to that of one Ch1 shell (see Supplementary Figure 10a).

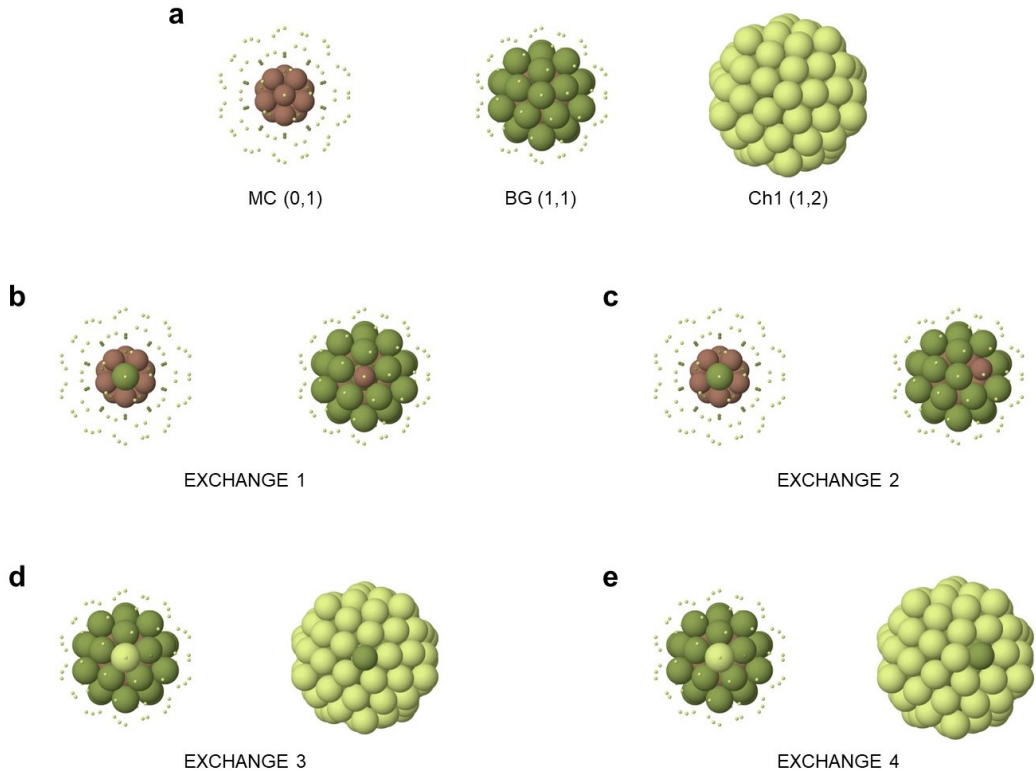

**Figure 13: Clusters of the DFT calculations.**
The structures considered in the calculations are of two types, with either $i = 3$ or $i = 4$ shells. In this figure we show only those containing BG and Ch1 shells, while the analogous structures containing 3 or 4 MC shells are not shown. The first structure ($i = 3$) contains the central atom $(0,0)$, the $(0,1)$ MC shell, and the $(1,1)$ BG shell, corresponding to the left and middle panels in **a**, for a total of 45 atoms. The second structure ($i = 4$) contains in addition the $(1,2)$ Ch1 shell, as shown in the right panel of **a**, for a total of 117 atoms. In **b**-**e** the different types of atom pair swaps are shown: in **b** and **c** between atoms of the $(0,1)$ and $(1,1)$ shells; in **d** and **e** between atoms of the $(1,1)$ and $(1,2)$ shells.

## S2 Density Functional Theory (DFT) calculations

In Supplementary Tables 2-4 we report the data for the atomic pair exchanges shown in Supplementary Figure 13 for different types of alkali metal clusters. DFT [51] and Gupta potential [52] results are reported. The results have been obtained as explained in the Methods section of the main text. Here we recall that the energy differences between the configuration after and before the exchange are related to locally minimized configurations.

All structures contain a core made of $(0, 0)$ and $(0, 1)$ MC shells, upon which further shells

58

are added, either of Ch1 class (the $(1,1)$ BG shell and the $(1,2)$ chiral shell) or of Ch0 class (the $(0,2)$ and $(0,3)$ MC shells). In the latter case, standard Mackay icosahedra are obtained. According to the estimates of the optimal mismatch values, for Na@K, Na@Rb and Na@K@Rb we expect that the clusters with Ch1 shells are more stable than the Mackay clusters. This is confirmed by the data reported in Supplementary Tables 2-4, where it is shown that Mackay clusters are energetically unstable upon atomic pair exchanges, while the clusters with Ch1 shells are stable. We note that DFT and Gupta results are in good agreement, which supports the validity of the Gupta model for these clusters.

|      | 3 SHELLS | | | | 4 SHELLS | | | |
|------|----------|--|--|--|----------|--|--|--|
| EXCH | MC@BG | | MC@MC | | MC@(BG-Ch1) | | MC@(MC-MC) | |
|      | DFT | GUPTA | DFT | GUPTA | DFT | GUPTA | DFT | GUPTA |
| 1 | +0.240 | +0.322 | −0.345 | −0.086 | +0.164 | +0.251 | −0.046 | +0.013 |
| 2 | +0.223 | +0.262 | −0.323 | −0.079 | +0.253 | +0.273 | −0.064 | +0.006 |

**Table 2: DFT and Gupta data for NaK clusters.**
In the initial configurations, these clusters always contain a core with MC shells $(0,0)$ and $(0,1)$ made of Na atoms, while the other shells are made of K atoms. For MC@BG, the $(1,1)$ BG shell is added to the core, whereas for MC@BG-Ch1 a further $(1,2)$ shell is added (see Supplementary Figure 13a). For MC@MC, the $(0,2)$ shell is added to the core, whereas for MC@MC-MC a further $(0,3)$ shell is added. Exchanges of type 1 and 2 (see Supplementary Figure 13b,c) are made on the initial configuration. The energy differences (in eV) between the configurations after and before the exchange of the atomic pair are reported. The exchange is favourable for negative differences, unfavourable otherwise.

|      | 3 SHELLS | | | | 4 SHELLS | | | |
|------|----------|--|--|--|----------|--|--|--|
| EXCH | MC@BG | | MC@MC | | MC@(BG-Ch1) | | MC@(MC-MC) | |
|      | DFT | GUPTA | DFT | GUPTA | DFT | GUPTA | DFT | GUPTA |
| 1 | +0.283 | +0.381 | −0.711 | −0.415 | +0.205 | +0.284 | −0.081 | −0.024 |
| 2 | +0.237 | +0.289 | −0.691 | −0.395 | +0.293 | +0.289 | −0.137 | −0.045 |

**Table 3: DFT and Gupta data for NaRb clusters.**
The same as in Table 2 but with Rb atoms instead of K atoms. Energies are in eV.

| EXCH | MC@BG@Ch1 | | MC@MC@MC | |
|---|---|---|---|---|
| | DFT | GUPTA | DFT | GUPTA |
| 1 | +0.180 | +0.223 | −0.025 | +0.031 |
| 2 | +0.242 | +0.237 | −0.042 | −0.006 |
| 3 | +0.064 | +0.071 | +0.031 | +0.047 |
| 4 | +0.048 | +0.054 | +0.019 | +0.028 |

**Table 4: DFT and Gupta data for NaKRb clusters.**
In the initial configurations, these clusters always contain a core with MC shells $(0,0)$ and $(0,1)$ made of Na atoms, while the third and fourth shells are made of K and Rb atoms, respectively. For MC@BG@Ch1, the $(1,1)$ and the $(1,2)$ shells are added to the core (see Supplementary Figure 13a), whereas for MC@MC@MC the $(0,2)$ and $(0,3)$ shells are added. Exchanges of types 1, 2, 3 and 4 (see Supplementary Figure 13b,e) are made on the initial configuration. The energy differences (in eV) between the configurations after and before the exchange of the atomic pair are reported. The exchange is favourable for negative differences, unfavourable otherwise.

## S3  Growth simulations

Here we present additional results on the simulations of the growth after deposition of atoms one-by-one on a preformed seed [24, 44]. These simulations show a few more examples of growth sequences that can be represented by paths in the hexagonal lattice.

In Supplementary Figure 14, three growth sequences for binary and ternary alkali metal clusters are shown. In Supplementary Figure 14a, we report the deposition of K atoms on a Na@K cluster terminated by a BG $(1,1)$ shell of K atoms. This is a case of deposition without mismatch, therefore the growth continues within the Ch1 class as expected.

In Supplementary Figure 14b, we show the result of the deposition of Cs atoms on a Na@Rb seed. The seed is terminated by a Ch1 $(1,3)$ shell of Rb atoms. According to the path rules, either a $(1,4)$ or a $(2,3)$ shell can form upon deposition of further atoms. The optimal size mismatch of these steps can be estimated by Eq. S53, where the coefficient $\xi$ is calculated recursively by applying Eq. S67 twice, since there are three shells made of the same atom type ($(1,1)$, $(1,2)$ and $(1,3)$). The optimal values of the mismatch are 0.0547 and 0.1162 for the steps towards $(1,4)$ and $(2,3)$, respectively. As usual, the class-conserving step is associated with the lower optimal mismatch. The mismatch between Cs and Rb, as evaluated from their bulk lattice constant, is of 0.077. Since Cs atoms are too large for accommodating on the $(1,4)$ shell, we expect the $(2,3)$ shell to form. This prediction finds confirmation in our growth simulations, where we observe the formation of a perfect Ch2 shell. We note that Cs and Rb atoms show some tendency towards intermixing [53], so that there are a few exchanges of Rb and Cs atoms during growth that bring some Rb atoms to the cluster surface.

In Supplementary Figure 14c a more complex growth sequence is shown. The seed is Na@K terminated by a Ch1 $(1,2)$ shell of K atoms, on which Rb atoms are deposited. Two steps are possible, leading to the formation of either the Ch1 $(1,3)$ shell or the BG $(2,2)$ shell. The

61

optimal values of the mismatch are 0.0618 and 0.1129, respectively. The mismatch between K and Rb is 0.0716, which is slightly larger than the optimal mismatch for the growth of a Ch1 shell. Indeed, as in the case of Supplementary Figure 14b, the chirality class is changed and the BG $(2,2)$ shell is formed. This is followed then by a $(2,3)$ Ch2 shell, i.e. the only viable alternative for the growth on top of the BG shell. However this is not the end of the story, because upon further Rb atom deposition there is first a sudden transformation of the subsurface shell from $(2,2)$ to $(1,3)$ and then of the $(2,3)$ shell from $(2,3)$ to $(1,4)$. These transformations take place by sudden collective reshaping processes, of the type frequently observed in metal clusters [24]. The final result is as if the cluster would have grown into the Ch1 class from the beginning. The transformation of the first Rb shell from $(2,2)$ to $(1,3)$ is due to the packing of multiple shells, which affects the optimal mismatch at the Na-Rb interface. Specifically, the addition of further shells slightly increases the optimal mismatch (see Supplementary Note 1.7). The optimal mismatch for $(1,2)$-$(1,3)$ interface is therefore likely to approach and eventually overcome the mismatch between K and Rb, so that the Ch1 shell becomes more favourable than the BG one. This induces the reshaping of the shell. After the transformation of the first Rb shell, we have a $(2,3)$ Rb shell on top of a $(1,3)$ shell made of atoms of the same type. This configuration is not the optimal one, because when the mismatch is zero atoms are better accommodated on shells of the same chirality class. Therefore, the transformation of the second Rb shell towards the Ch1 $(1,4)$ arrangement takes place.

In Supplementary Figure 15 we show a growth sequence for a alkali metal cluster made of four different elements. The seed is Na@K@Rb terminated by a Ch1 $(1,3)$ shell of Rb atoms, on which Cs atoms are deposited. Two steps are possible, leading to the formation of either the Ch1 $(1,4)$ shell or the Ch2 $(2,3)$ shell. The optimal values of the mismatch are 0.0183 and 0.0778, respectively. We note that these values are different from those of Supplementary Figure 14, referring to the same steps on the closed-packed plane, because here in the seed

there is a single shell of Rb atoms instead of three. The mismatch between Rb and Cs is 0.0769, which is very close to the optimal mismatch for the formation of the Ch2 $(2,3)$ shell. Such shell is indeed formed in our simulations. However, the deposition of further Cs atoms leads to the formation of the BG $(3,3)$ shell instead of the Ch2 $(2,4)$ shell, which is what one would expect when atoms of the same species are deposited. However, we note that, in this case, the optimal mismatch for both steps is quite small, being 0.0249 for the class-conserving step and 0.0459 for the class-changing one. As such, even though the Ch2 shell is expected to be more favourable, the BG one is not that unfavourable and therefore may form during the growth simulation. Besides, due to the mixing of Cs atoms with the smaller Rb atoms in the $(2,3)$ shell, the effective mismatch is actually larger then zero, which helps in stabilizing the BG shell. Upon further deposition of Cs atoms, the Ch3 $(3,4)$ shell is formed, corresponding to the natural growth step on top of the BG $(3,3)$ shell.

In Supplementary Figure 16 data on the growth of Ag atoms on Ni and Co Mackay seeds of 147 and 309 atoms are shown. The large mismatch (0.161 for Ag on Ni and 0.156 for Ag on Co) is close to optimal for the growth of Ch1 shells on seeds in this size range; this kind of growth is indeed observed in the growth simulations.

In Supplementary Figure 17 the growth of Ag atoms on a Mackay seeds of size 309 is shown. This cluster is terminated by a $(0,4)$ shell. For the growth of the first Ag shell, the mismatch between Ag and Cu (0.129) is closer to the optimal one of an AM1 shell. The AM1 $(1,4)^*$ shell is indeed grown, with very few defects. Then the growth on the AM1 shell continues by a class-changing path by forming defective AM shells up to the AM3 $(1,4)^*$ shell. Upon further deposition, there is a sudden transformation of all Ag shells to the Ch1 arrangement, up to the $(1,6)$; this configuration is energetically more favourable.

In Supplementary Figure 18 the growth of Ag on imperfect Mackay Ni seeds is shown. These seeds present an incomplete Mackay external shell. In spite of the defective character of

the seed, Ch1 shells are formed. These shell are slightly distorted because of the asymmetries in the starting core, but they very much resemble the perfect Ch1 shells. This shows that the growth mode of chiral shell is quite robust against imperfections.

In Supplementary Figure 19 the growth mechanism of AM shells one on top of the other is discussed. This mechanism is dominated by the trapping of atoms in fourfold adsorption sites on the surface, which naturally leads to class changes at every step in the hexagonal plane.

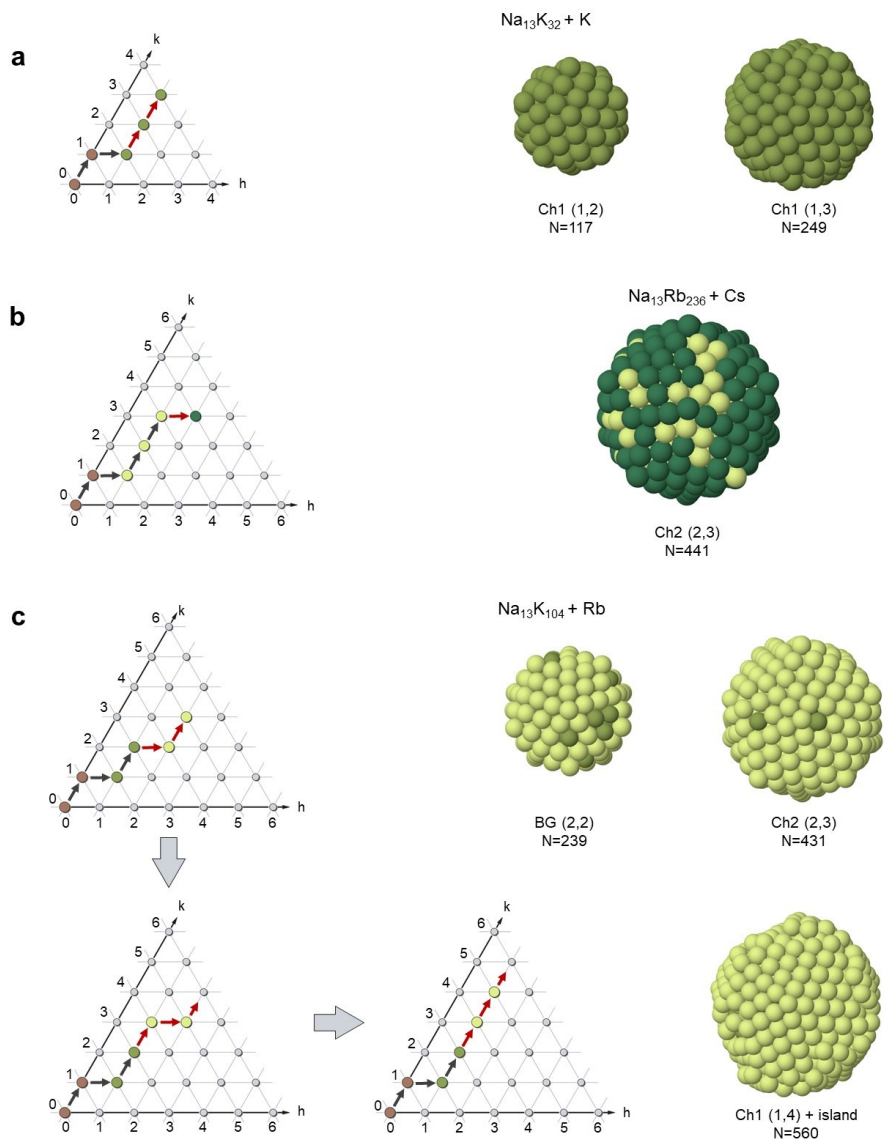

**Figure 14: Growth simulations of alkali metal clusters of two and three different elements.**
**a** Deposition of K atoms on a Na@K seed of 45 atoms, in which a Mackay core of 13 atoms is covered by one Ch1 shell of K atoms. The growth temperature is 125 K and the deposition rate is 1 atoms/ns. Deposited K atoms form two further Ch1 shells. **b** Deposition of Cs atoms on a Na@Rb seed of 249 atoms, in which a Mackay Na inner core is covered by two Ch1 shells of Rb atoms. The growth temperature is 125 K and the deposition rate is 1 atoms/ns. Cs atoms form a Ch2 shell. **c** Deposition of Rb atoms on a Na@K seed of 117 atoms, made of a Mackay inner core of 13 Na atoms and two Ch1 shells of K atoms. Growth temperature is 125 K and deposition rate is 1 atoms/ns. Deposited Rb atoms first form the $(2,2)$ BG shell, which is completed at 239 atoms, and then the growth continues by the formation of a $(2,3)$ Ch2 shell. Upon deposition of further Rb atoms, the $(2,2)$ shell suddenly transforms into the $(1,3)$ Ch1 shell, still covered by a $(2,3)$ Ch2 shell, which then transforms into the $(1,4)$ Ch1 shell.

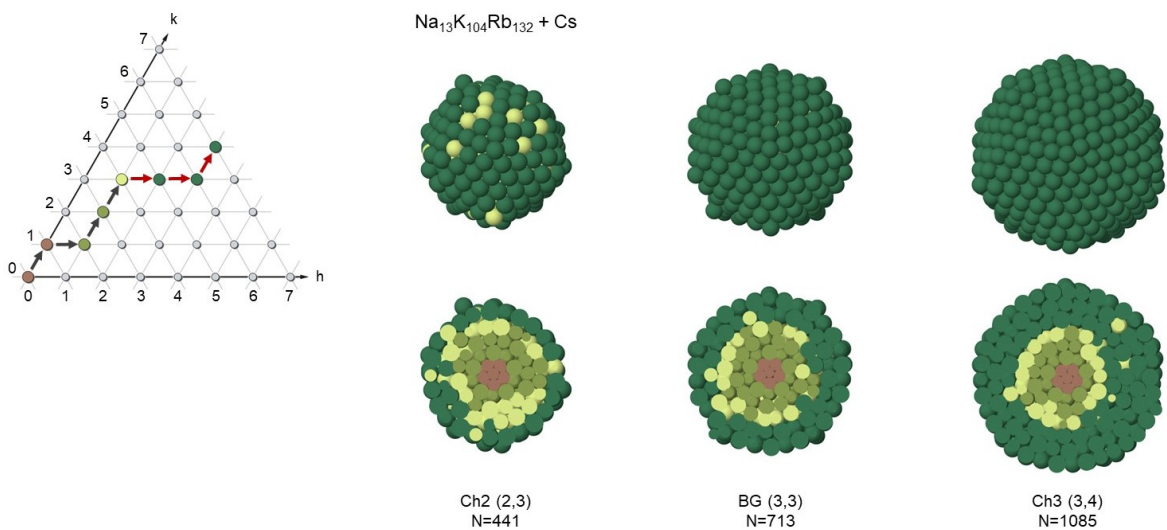

**Figure 15: Growth simulations of alkali metal clusters of four different elements.**
Deposition of Cs atoms on a Na@K@Rb seed of 249 atoms, in which a Mackay core of 13 atoms is covered by two Ch1 shell of K atoms, and by a Ch1 shell of Rb. The growth temperature is 125 K and the deposition rate is 1 atoms/ns. Deposited Cs atoms form Ch2 and Ch3 shells

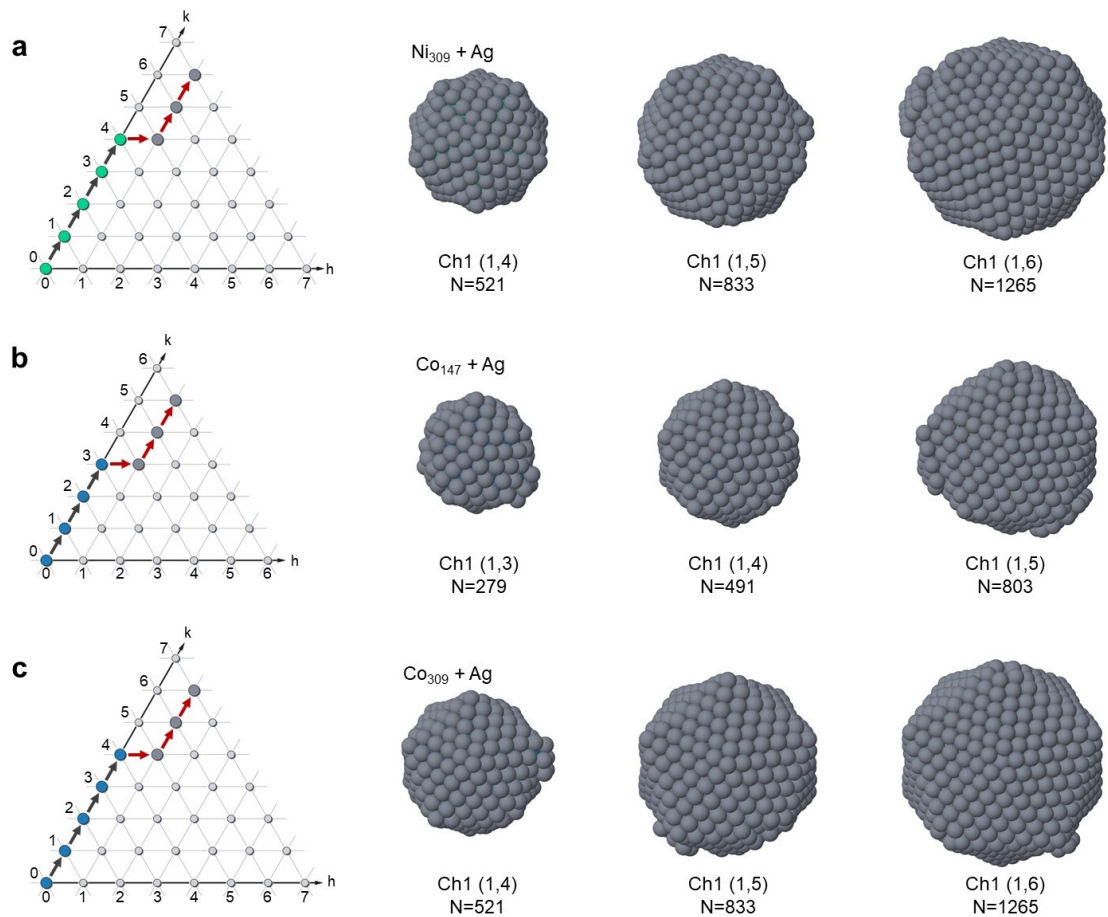

**Figure 16: Growth sequences for Ag atoms on Ni and on Co seeds.**
**A** Deposition of Ag atoms on a Ni$_{309}$ Mackay icosahedral seed, at $T = 450$ K, and deposition rate 0.1 atoms/ns.
**B** Deposition of Ag atoms on a Co$_{147}$ Mackay icosahedral seed, at $T = 350$ K, and deposition rate 0.1 atoms/ns.
**c** Deposition of Ag atoms on a Co$_{309}$ Mackay icosahedral seed, at $T = 450$ K, and deposition rate 0.1 atom/ns. In **A-C**, we show the growth path (the part of the path corresponding to the red arrows) in the hexagonal lattice and images of the cluster surface at different growth stages.

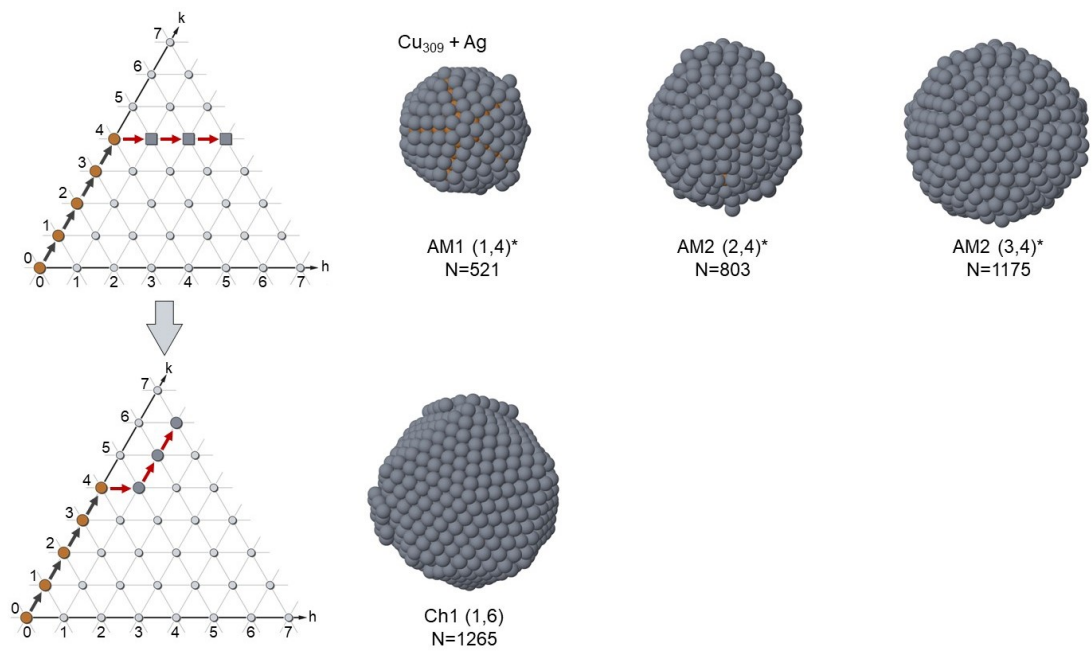

**Figure 17: Growth sequence for Ag atoms on a Cu seed.**
In the top row, the growth sequence is shown for the deposition of Ag atoms on a $Cu_{309}$ Mackay icosahedral seed, at $T = 350$ K, and deposition rate 1 atom/ns up to size 1175. The growth produces AM structures changing class at each step in the hexagonal plane. The bottom row show the continuation of the growth up to size 1265. In this part of the simulation, a sudden rearrangement of the Ag shells to Ch1 structures takes place, so that the final structure does not contain AM shells anymore and its structure is represented by the the path in the bottom row. The sudden rearrangement is due to the very unfavourable energy of the AM multi-shell arrangement, which is highly metastable.

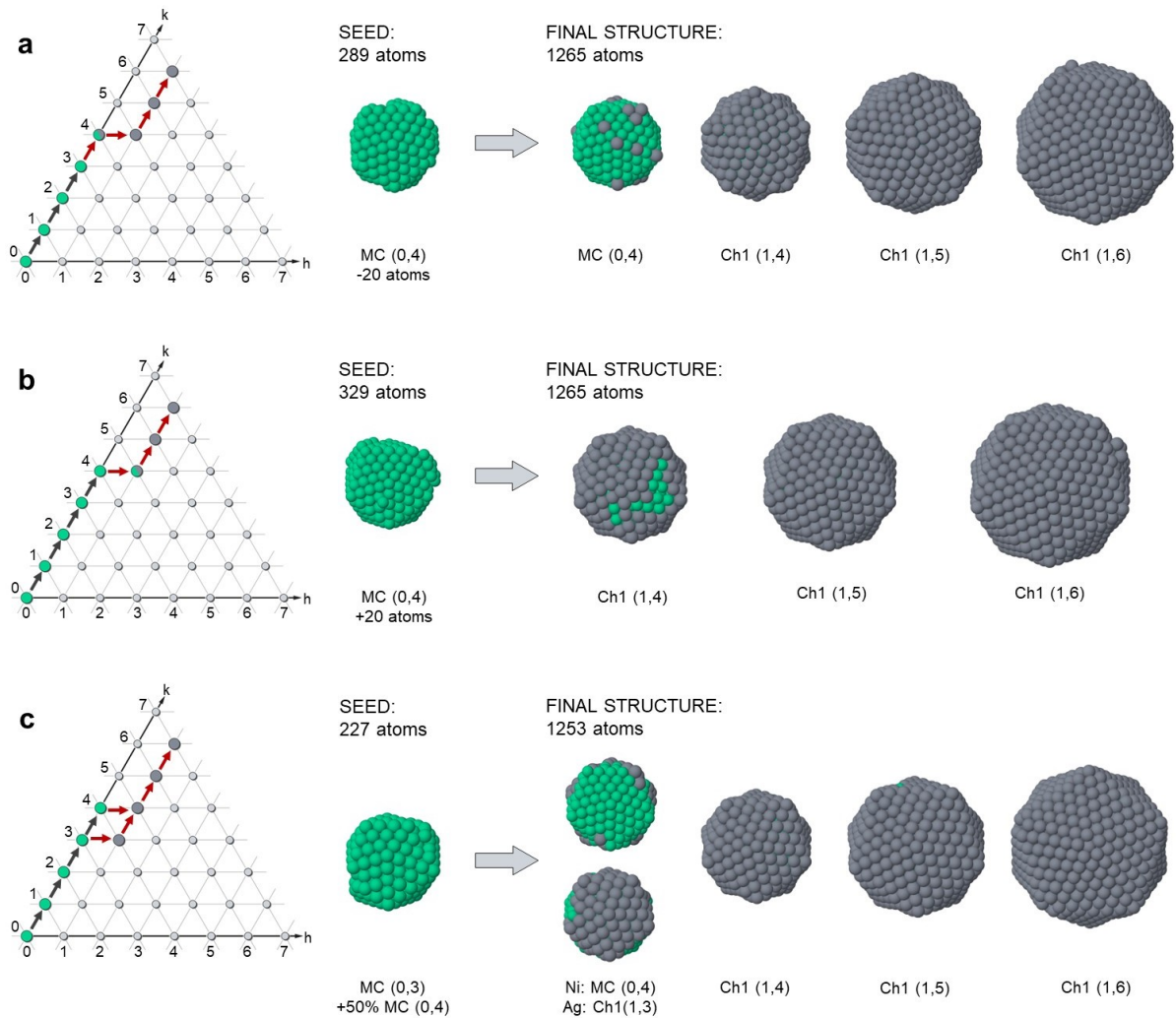

**Figure 18: Growth of Ag atoms on a Ni defective seeds.**
The growth simulations are at $T = 450$ K, and deposition rate 0.1 atoms/ns. At variance with Supplementary Figures 16 and 17, here the images represent different shells of the final structures, obtained at the end of the growth, instead of snapshots taken during the growth. In **a** growth starts from a $Ni_{289}$ seed, which is terminated by a slightly incomplete $(0, 4)$ MC shell. In the final structure, the $(0, 4)$ shell is completed by Ag atoms, then covered by Ch1 shells. In **b** growth starts from a $Ni_{329}$ seed, which is terminated by a small fragment of a $(0,5)$ MC shell. in the final structure, that shell is completed by Ag atoms, which transform it into a Ch1 $(1, 4)$ shell, which is covered by further Ch1 shells. In **c** the growth starts from a $Ni_{227}$ seed, which is terminated by a half $(0, 4)$ MC shell. In the final structure, that shell is completed by Ag atoms, so that a the resulting shell presents a Ni and an Ag half. On the Ni side, the shell is MC $(0, 4)$, while on the Ag side the shell is Ch1 $(1, 3)$. This is indicated by the bifurcation in the path. Further shells are of class Ch1.

69

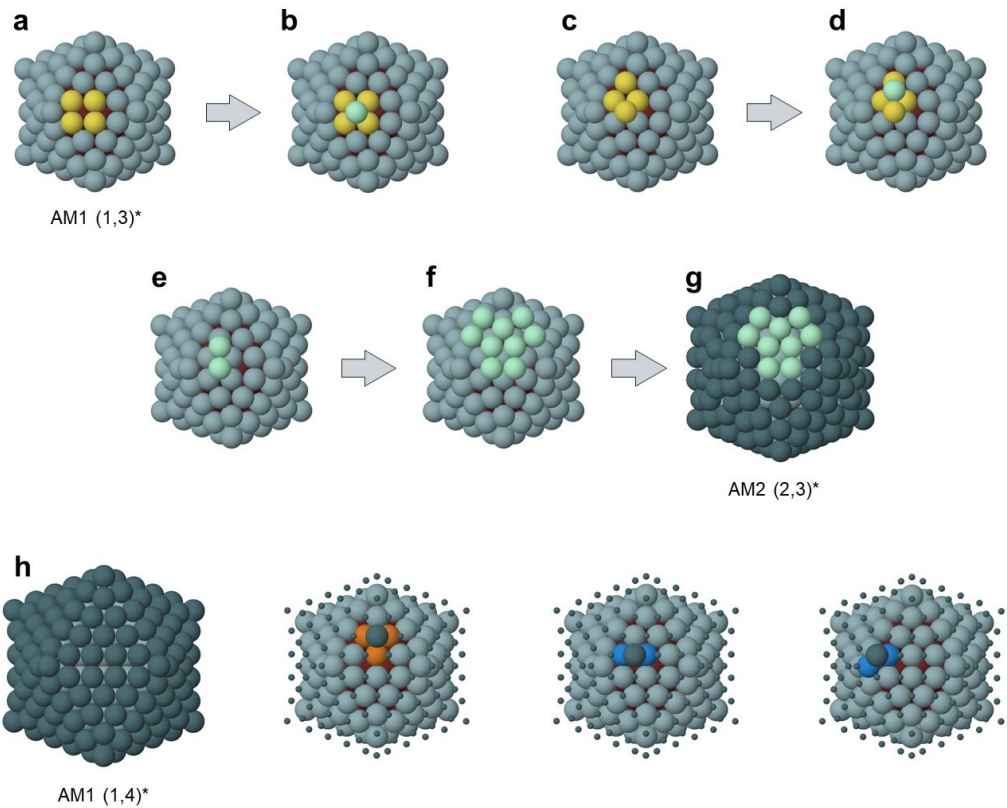

**Figure 19: Growth mechanism on AM shells.**
**a** The four orange atoms identify a fourfold adsorption site on the $(1, 3)^*$ AM shell. **b** A deposited adatom (cyan) is preferentially trapped in a site of this type. **c** A new fourfold adsorption site is created on top for the four orange atom so that **d** it can trap a further deposited atom. **e-g** The process is self-replicating, finally leading to the formation of the $(2, 3)^*$ AM shell. **h** The atoms of $(1, 4)^*$ AM shell on top of a $(1, 3)^*$ AM shell find only threefold adsorption sites, identified by the orange atoms. Sites at the facet edge are twofold (see the blue atoms) so that they are not even locally stable for isolated adatoms.



\end{document}